\documentclass[vecphys]{svmult}
\pdfoutput=1
\usepackage[dvips]{color}
\usepackage{makeidx}         
\usepackage{graphicx}        
\usepackage{multicol}        
\usepackage{cite}            
\usepackage[bottom]{footmisc}
\usepackage{amsmath}
\usepackage{subfigure}
\usepackage{tikz}		
\usepackage{floatrow}
\makeindex             

\begin{document}
\title{Single-Chain Magnets}
\author{$^{1}$Dante Gatteschi \and $^{2}$Alessandro Vindigni}
%
\institute{Department of Chemistry, 
University of Florence and INSTM
\vspace{0.2cm} 
\and
Laboratory for Solid State Physics, 
ETH Zurich
}

\maketitle

\begin{abstract} 
Single-chain magnets are molecular spin chains displaying slow relaxation of the magnetisation on a macroscopic time scale. 
To this similarity with single-molecule magnets they own their name. In this chapter the distinctive features of single-chain magnets 
as opposed to their precursors will be pinpointed. In particular, we will show how their behaviour is dictated by the physics of thermally-excited domain walls. 
The basic concepts needed to understand and model single-chain magnets will also be reviewed. 
\end{abstract}

\noindent

\section{Introduction\label{intro}}
The observation of magnetic hysteresis of molecular origin in Single-Molecule Magnets (SMMs) is considered one of the most relevant 
achievements in nanomagnetism~\cite{Gatteschi-Sessoli_AC_03,G-S-V}.  
Fundamental aspects related to quantum tunnelling of the magnetisation have been thoroughly discussed in the previous chapters. 
On a more practical perspective, that observation rendered the molecular approach 
one of the possible routes to realizing bistable nano-objects, suitable for magnetic storage or quantum-computing applications.    
In spite of many efforts, the highest blocking temperature attained by SMMs remains, still nowadays, in the liquid-helium temperature range.    
The idea that one-dimensional (1D) structures of coupled paramagnetic ions might afford higher blocking temperatures started developing at the end of the nineties and the first examples of slowly relaxing 1D systems were reported at the beginning of the new century~\cite{Angew,Coulon_PRB_04}.  
The resulting molecular systems have been dabbed Single-Chain Magnets (SCMs) in order to evidence analogies with their precursors, SMMs, 
while remarking -- at the same time -- the 1D character.  
In some cases, SMMs themselves have been employed as building blocks for such 1D magnetic lattices~\cite{JACS_Lecren}. 
With the aim of increasing the blocking temperature as much as possible, different synthesis strategies have been followed to obtain 
some type of magnetic anisotropy at the level of building blocks or of the coupling among them. 
In the present chapter we will be dealing with uniaxial anisotropies only, though this requirement is not strict for the observation of 
SCM behaviour~\cite{Verdaguer_review,Review_Gao,Yamashita_RSC_2012}. 

A distinctive feature of 1D magnetic systems is the development of short-ranged correlations upon cooling. 
This makes them substantially different from both paramagnets and bulk magnets. 
Should one establish an analogy between classical magnetic ordering and phases of matter, paramagnets would be identified with perfect gases  
while bulk magnets with solids. Pushing this naive analogy further, spin chains would be associated with liquids, 
specifically in the temperature range in which short-range correlations extend over several lattice units. 
The degree of spatial correlation is quantified by the  correlation length. 
In molecular chains consisting of magnetic building bocks with uniaxial anisotropy, the correlation length 
typically diverges exponentially  with decreasing temperature. 
From a snapshot taken at finite temperature, any chain would appear as a collection of randomly oriented magnetic domains%
\footnote{These soft, fluctuating domains should not be confused with Weiss domains encountered in magnetically ordered phases.}  
separated by domain walls (DWs). 
The average size of those domains is of the order of the correlation length. 
This pictorial, but essentially correct, scenario is consistent with thermally-driven diffusion of DWs.   
In this sense, the response of a SCM to a tiny a.c. field is expected to be determined by the time needed to adjust the size of domains to the external stimulus.     
A robust scaling argument associates the characteristic time of this readjustment with the time elapsed while a DW diffuses 
over a distance proportional to the correlation length. Within this idealized picture, the relaxation time of the magnetisation is expected to scale with temperature 
like the square of the correlation length. 

The qualitative description given above applies to the ideal case of infinite chains and small applied fields. 
The first hypothesis is practically never fulfilled in real systems. In fact, the number of magnetic centres interacting consecutively is typically limited to 10$^2$--10$^4$     
by naturally occurring defects, non-magnetic impurities or lattice dislocations~\cite{PRL_Bogani,APL_Vindigni,Gambardella}.  
A SCM may thus behave as if it extended indefinitely only when the correlation length is much smaller than the average distance between successive defects. 
Upon lowering the temperature, a crossover is expected at which the correlation length 
becomes of the order of the average distance among defects. 
Below this crossover temperature, spins enclosed between two successive defects are parallel with each other and no DW is 
present at equilibrium. In this \textit{finite-size regime} relaxation is somewhat equivalent to magnetisation reversal in 
nanoparticles or nanowires, which may occur via N\'eel-Brown uniform rotation or by droplet-nucleation mechanism~\cite{B_Braun_AdvPhys_2012}.
    
All the mentioned mechanisms represent  potential channels for relaxation in SCMs. 
Which one is favoured depends on the experimental conditions: temperature, applied field and amount of defects in the sample. 
The random-walk argument which relates the correlation length to the relaxation time holds in the linear-response regime, i.e., 
when such tiny fields are applied to induce just slight deviations from thermodynamic equilibrium.  
On the contrary, relaxation from a saturated configuration typically entails far-from-equilibrium dynamics.
In this type of experiments nucleation of soliton-antisoliton pairs or of a single DW adjacent to a defect possibly initiates the relaxation process. 
N\'eel-Brown uniform rotation practically represent an alternative channel for relaxation only for very short segments of chain, encountered in samples 
in which finite-size effects have been enhanced by doping with non-magnetic impurities~\cite{PRL_Bogani,APL_Vindigni}. 
   
It should not be forgotten that molecular spin chains are packed in three-dimensional crystals.  
Though several synthesis strategies may be followed to minimize interactions among chains, at least the dipolar interaction cannot be suppressed completely. 
Therefore, below some temperature, a 3D magnetically ordered phase is expected to appear.   
Whether such a phase is observed or not in a specific compound depends on how long 
the relaxation time is at the transition temperature~\cite{Coulon_PRL_09}. 
Generally, when the time needed for the system to equilibrate is much longer than experimental time scales,
the distinctive features of the underlying equilibrium phase, possibly ordered, cannot be evidenced.  
For weakly interacting spin chains, the expected transition temperature is much higher than interchain interaction in Kelvin units ($kT_\text{C}\gg J'$ with the forthcoming notation). 
In fact, the 3D-ordering process is ``assisted'' by the development of strong short-range correlations 
inside each chain~\cite{Scalapino_quasi_1d,Kai_IC_89,Ferrero_JACS_91}. 
However, in realistic samples, defects prevent the intrachain  correlation length from diverging indefinitely, which eventually lowers the transition temperature 
to the ordered phase.    
In several SCMs slow dynamics was observed down to few Kelvins, before 3D ordering took place, right because of the presence of defects and non-magnetic impurities.   

Both SMMs and SCMs are characterized by slow dynamics of molecular origin, acting at macroscopic time scales and in the absence of 3D magnetic ordering.
Even if impurities play a crucial role in SCMs, they usually do not bring enough disorder to give rise to spin-glass behaviour.   
Consistently, slow dynamics is typically characterized by a single time scale which does not display a super-Arrhenius behaviour at any temperature~\cite{Glassy-reviews}. 
Besides preventing the onset of 3D magnetic ordering, the increase of relaxation time with cooling usually leads to complete blocking before genuine quantum effects become evident~\cite{Coldea}. 

From what written till now, it should be clear that many effects may interplay in determining the magnetic behaviour of spin chains. 
We will focus on those systems in which slow dynamics can be ascribed to each single chain and does not originate from cooperative 3D interactions. 

The goal of this chapter is that of highlighting the properties of SCMs with a critical view to what has been done and what still deserves further investigation. We will not try to cover in detail all the representative literature, for which the reader is addressed to excellent reviews~\cite{Review_Gao,Review_CC,Verdaguer_review,Yamashita_RSC_2012}. Although SCMs have been widely investigated, the interest of the physics community has not been comparable to that shown for SMMs. This is partially due to the fact that the novelty of SCMs compared to traditional 1D spin systems hardly emerged. Here we attempt to provide and efficient overview of the essential, novel physics of SCMs and hope that the final comment will be more benevolent. 
The chapter is organised as follows: Sect. 1.2 will cover the basic aspects of classical spin chains; the chemical frame will be discussed within a bottom-up or building-block approach in Sect. 1.3; in Sect. 1.4 the spin Hamiltonians typically used to rationalise the physical properties of SCMs will be introduced; relevant extensions of the Glauber model developed in the SCM context without and with defects will be treated in Sect. 1.5 and 1.6, respectively; phenomenological arguments not contained in the Glauber model but relevant for understanding SCMs  will be discussed in Sect. 1.7; a short section on perspectives will conclude the chapter.

\section{Thermal equilibrium and slow dynamics in ideal SCMs\label{origin_of_slow_dyn}}
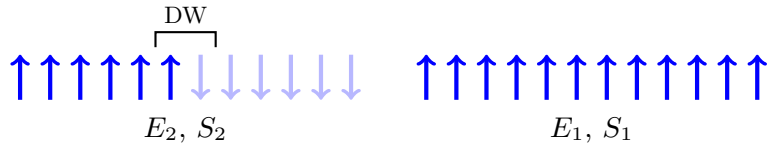
\begin{figure}[ht]
\vspace{-0.4cm}
\begin{center}
\begin{tikzpicture}[scale=1, ultra thick]
	\foreach \p in {0.2,0.6,...,2.21}
 	\draw [<-,blue] (\p,0.3) -- (\p,-0.3);
	\foreach \p in {2.6,3,...,4.6}
 	\draw [->,blue!28!white] (\p,0.3) -- (\p,-0.3);
	\foreach \p in {5.6,6,...,10}
 	\draw [<-,blue] (\p,0.3) -- (\p,-0.3);
	\draw [thin,white](5.6,-0.4) -- (10,-0.4) node [below,midway,scale=1.2] {$\textcolor{black}{E_1, \,S_1}$};
	\draw [thin,white](0.2,-0.4) -- (4.6,-0.4) node [below,midway,scale=1.2] {$\textcolor{black}{E_2, \,S_2}$};
	\draw [thick](2.,0.4) -- (2,0.6) -- (2.8,0.6)  node [above, midway] {\textcolor{black}{DW}} -- (2.8,0.4);
\end{tikzpicture}
\vspace{-0.4cm}
\caption{Sketch representing the configurations whose free-energy difference is evaluated in the 
text: a ferromagnetic ground state with all the spins parallel to each other (right) and a configuration consisting of two domains with opposite
spin alignment (left). \label{Sketch_Ising_DW}}
\end{center}
\end{figure}

In this section the peculiarities of classical spin chains with uniaxial anisotropy that directly affect the physics of SCMs will be recalled.  
Indeed, the distinctive feature of SCMs is that of approaching thermodynamic equilibrium slowly. By \textit{slowly} we mean   
that relaxation time becomes longer than milliseconds at temperatures of the order of 10 K or lower. The reference equilibrium state to be reached is also relevant.   
As already mentioned, as long as 3D interactions are negligible, no magnetisation is expected in zero applied field at thermodynamic equilibrium. 
Long-range magnetic order may be destroyed by thermally-excited spin waves or DWs either. The first ones are effective in the absence of anisotropy, 
according to the Mermin--Wagner theorem~\cite{Mermin-Wagner,Fisher}.   
The fact that disordering is, instead, driven by DWs in the presence of anisotropy
can be easily understood recalling an argument presented in the Landau--Lifshitz series~\cite{Landau_book_stat_phys}. 
Let us consider a group of $N$ spins that preferentially point along the same direction, say up or down. 
For the moment we assume the axes of easy anisotropy to be collinear, as represented schematically in Fig.~\ref{Sketch_Ising_DW}.
We evaluate the variation of the free energy associated with the creation of a DW
starting from a configuration with all the spins parallel to each other.
Creating a DW increases the energy by a factor $E_2-E_1=\mathcal{E}_\text{dw}$. 
On the other hand, such a DW may occupy $N$ different positions in the spin chain, so that
the relative entropy increase scales as $S_2-S_1 =k \ln(N)$. The free-energy difference
between the two configurations sketched in  Fig.~\ref{Sketch_Ising_DW} is roughly 
$\Delta F \simeq \mathcal{E}_\text{dw} - k T\ln (N)$. 
When the thermodynamic limit $N\rightarrow\infty$ is taken, one immediately realizes that it is always convenient to split
the system into groups of parallel spins. As a consequence, long-range magnetic order is destroyed at any finite temperature.  
In principle, in an infinite chain, the same mechanism may allow creating an indefinite number of DWs.  
However, the average distance among them does depend on temperature and it is inversely proportional to the  correlation length~\cite{Schriffer}.  
It is worth remarking that in the text-book argument given above the following assumptions have been implicitly made:
\begin{enumerate} 
\item that DWs extended only over one lattice unit
\item spin-wave excitations were not considered 
\item the thermodynamic limit was taken.  
\end{enumerate} 
Whether the first hypothesis is fulfilled or not depends on the relative strength of exchange interaction and magnetic-anisotropy energy. 
This can be discussed more concretely by considering the classical Heisenberg model with uniaxial anisotropy:
\begin{equation}
\label{Heisenberg_chain_classic}
\mathcal{H}_\text{H}= -\sum_{ i=1}^{N} 
\left[J\vec S_i\cdot\vec S_{i+1}+D\left(S_i^z\right)^2\right]\;,
\end{equation}
where $\vec S_i$ are classical spins, $J$ and $D$ the exchange 
and the magnetic-anisotropy energy, respectively; $\left| \vec  S_i  \right|=1$ will be assumed henceforth.  
Though it does not entail the complexity of many real systems, Hamiltonian~\eqref{Heisenberg_chain_classic} is 
a useful reference to discuss the physics of SCMs. 
To the aim of distinguishing between two simple types of DWs, we fix 
$D\!>\!0$ and $J\!>0$. 
With Hamiltonian~\eqref{Heisenberg_chain_classic}, DWs can be larger than one lattice spacing. 
In fact, the actual DW profile results from the \textit{competition} between the exchange energy (which is minimized by broadening the wall) 
and the anisotropy energy (which favours a sharp wall). 
Domain walls whose structure develop over more lattice units will be referred as \textit{broad}; these are opposed to \textit{sharp} DWs  
in which the local magnetisation changes abruptly its sign, within one lattice distance.    
The energy associated with a broad DW is $\mathcal{E}_\text{dw}=2\sqrt{2 DJ}$~\cite{Enz}, namely the energy needed to create one soliton ``particle'' in the spin chain~\cite{Steiner_Mikeska}. 
For sharp DWs, one obtains $\mathcal{E}_\text{dw}=2J$, as per the Ising model.  
The crossover from sharp- to broad-wall occurs at $D/J=2/3$~\cite{ICA_Vindigni_08,Barbara,Yan-Bauer_PRL_2012}.  
The analytic formula for broad-DW energy, $\mathcal{E}_\text{dw}=2\sqrt{2 DJ}$, was obtained in the continuum formalism 
and gets less and less accurate as the threshold ratio is approached from below, $D/J\rightarrow(2/3)^-$.

If the Landau's argument is rephrased for DW excitations of finite thickness $w\!=\!\sqrt{J/2D}$, 
the counting of equivalent configurations with the same energy needs to be modified and -- in turn -- 
the entropy contribution $S_2-S_1 =k \ln(N/w)$. 
In this case, splitting the uniform configuration into domains becomes convenient as soon as the number of spins 
exceeds the product $w \, \E^{\mathcal{E}_\text{dw}/kT}$. 
The latter threshold gives an estimate of the average number of consecutive spins that can be found aligned at a given temperature. 
To the leading order, the  correlation length scales in the same way at low temperature: $\xi \sim w \, \E^{\mathcal{E}_\text{dw}/kT}$.  
The energy $\mathcal{E}_\text{dw}$ represents the  natural ``unit'' which controls the divergence of the correlation length. 
Thus, in classical spin chains with uniaxial anisotropy 
the characteristic exponential divergence of $\xi$ is closely related to the fact that ferromagnetism is destroyed by thermally excited DWs.  
\begin{figure}[ht]
\vspace{-0.2cm}
\begin{center}
\includegraphics[width=11cm]{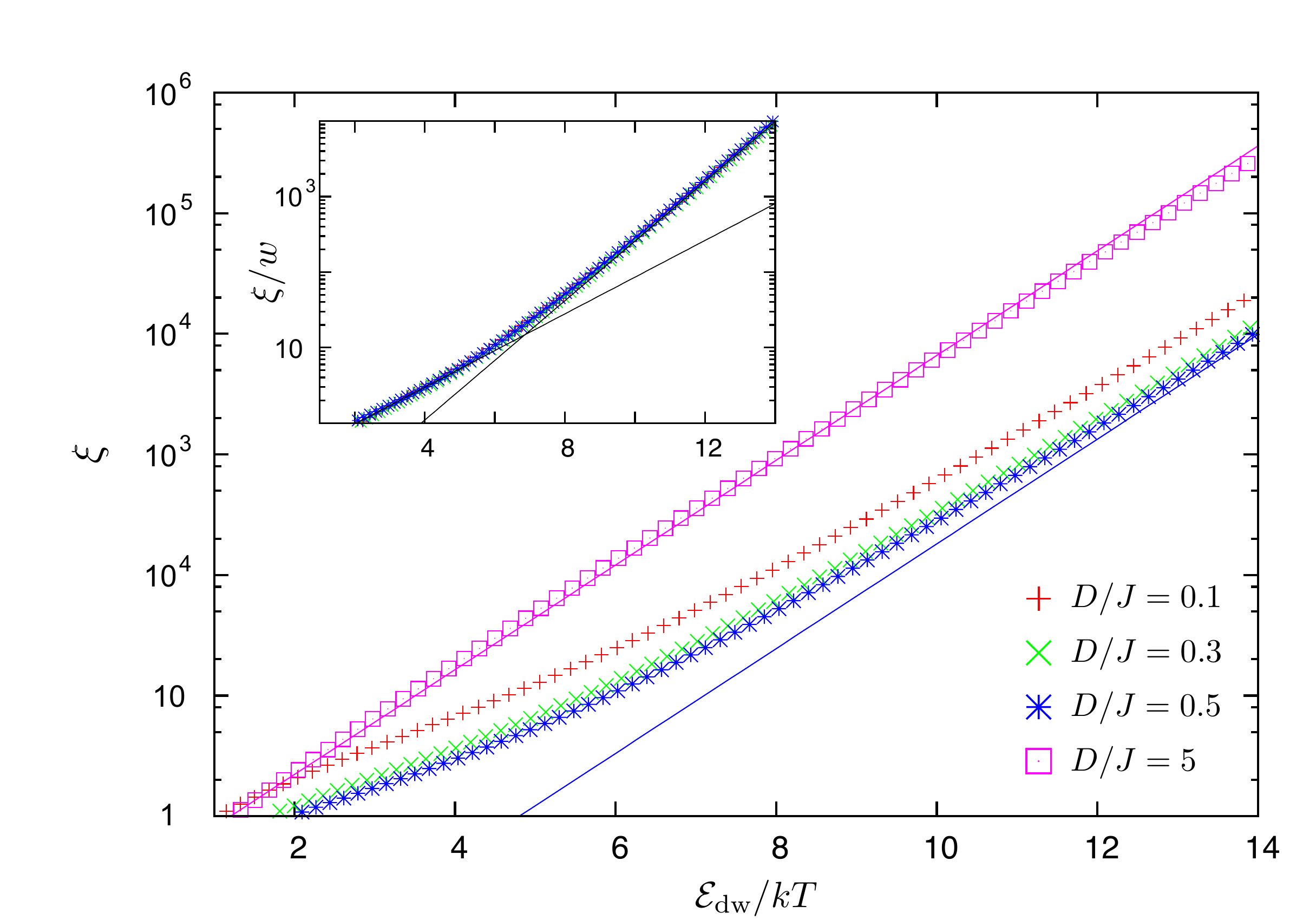}
\vspace{-0.4cm}
\caption{\label{fig_xi} 
Log-linear plot of $\xi$ in lattice units computed with the transfer-matrix technique as function of the ratio $\mathcal{E}_\text{dw}/kT$
for different values of $D/J$. 
For $D/J$= 0.1 (red crosses), 0.3 (green crosses), 0.5 (blue stars) DWs are \textit{broad} and $\mathcal{E}_\text{dw}$ has been computed 
numerically on a discrete lattice. For $D/J$=5 (open squares) DWs are \textit{sharp} and $\mathcal{E}_\text{dw}$=$2J$ has been used. 
The two solid lines give the ``reference'' behaviour $\xi\!\sim\! \E^{\Delta_\xi /kT}$ which is indeed followed when  
DWs are \textit{sharp} ($D/J$=5) but not when DWs  broaden. Inset: $\Lambda$=$\xi/w$ is plotted as a function of $\mathcal{E}_\text{dw}/kT$ for the values of $D/J$ consistent with \textit{broad} DWs. 
The universality of $\Lambda$ is highlighted by the data collapsing. Solid lines evidence the decrease of $\Delta_\xi$ with increasing temperature~\cite{Billoni_PRB}.  }
\end{center}
\end{figure}

In contrast to the Ising model~\cite{Ising}, the classical spin Hamiltonian~\eqref{Heisenberg_chain_classic} 
can also host spin-wave excitations, besides DWs.  
Due to the interaction between spin waves and broad DWs an additional temperature-dependent factor appears in front of 
the exponential in the low-temperature expansion of the correlation length~\cite{Fogedby_JPC_84}. 
Moreover, spin waves renormalise the DW energy at intermediate temperatures. 
The net result of the complicated interplay between thermalised spin waves and DWs is that the energy barrier controlling the divergence of $\xi$ 
(usually called $\Delta_\xi$ in SCM literature~\cite{Review_Gao,Review_CC,Verdaguer_review,Yamashita_RSC_2012}) is generally smaller than $\mathcal{E}_\text{dw}$ and takes different values   
depending on the temperature range in which it is measured~\cite{Billoni_PRB}. 
A similar effect was reported for the activation energy of 2$\pi$ sine-Gordon solitons in Mn$^{2+}$-radical spin chains~\cite{Ferrero_Mol_Phys_95}. 
Fig. \ref{fig_xi} highlights how $\Delta_\xi$ is constant and equal to $\mathcal{E}_\text{dw}$ for sharp DWs, while it varies significantly for broad DWs. 
However, the correlation length in units of $w$ keeps depending only the ratio $\mathcal{E}_\text{dw}/kT$, i.e., 
$\xi/w=\Lambda\left(\mathcal{E}_\text{dw}/kT\right)$. The inset shows that the curves corresponding to broad DWs indeed collapse onto each other when the ratio  
$\xi/w$ is plotted as a function of $\mathcal{E}_\text{dw}/kT$. 

As mentioned in the introduction, in realistic spin chains the divergence of the correlation length is always hindered 
by the presence of defects and non-magnetic impurities. 
This implies that results derived taking the thermodynamic limit, $N\rightarrow\infty$, do not 
hold down to indefinitely low temperatures.  
If we assume -- for the time being -- an idealized scenario in which such defects do not occur, a certain number of 
DWs shall be present at any finite temperature. A simple random-walk argument then relates the relaxation time to the correlation length: 
within a time $\tau$ a DW performs a random walk over a distance proportional to $\xi$~\cite{Tobochnik}. 
In other words, the relation 
\begin{equation}\label{xi-tau}
\xi^2\simeq 2D_\text{s}\tau
\end{equation}
\noindent
can be assumed, with $D_\text{s}$ being the diffusion coefficient. The latter generally increases with increasing temperature. Moreover,
it is expected to depend on temperature differently for sharp or broad diffusing DWs.  
When presenting the Glauber model we will see that $D_\text{s}$ can also be interpreted as the attempt frequency to flip a spin 
adjacent to a sharp DW.  

Summarizing, the presence of uniaxial anisotropy produces an exponential divergence of the correlation length with decreasing temperature. 
As the relaxation time is related to $\xi$ by a random-walk argument, it is also expected to diverge likewise, so that   
\begin{eqnarray}\label{activation_xi-tau}
\xi\sim \E^{\Delta_\xi/kT}\quad\quad\quad
\tau\sim \E^{\Delta_\tau/kT}\,.
\end{eqnarray}
In ideal 1D magnetic systems~\cite{Steiner_Mikeska,Villain} the correlation length is proportional to the product of 
temperature by static susceptibility (measured in zero field):  
\begin{equation}\label{xi_chi_stat}
\chi_\text{eq}\,T\sim \xi\;.
\end{equation}
The relaxation time can, instead, be obtained from dynamic susceptibility measurements as follows
\begin{equation}\label{chi_ac}
\chi(\omega, \,T) = \frac{\chi_\text{eq}}{1-i\omega\tau} \;,
\end{equation}
\noindent where $\omega$ is the frequency of the oscillating applied field and $\chi_\text{eq}$ is the static susceptibility%
\footnote{The more general Cole-Cole equation is needed when relaxation is not characterised by a single $\tau$ or to account for adiabatic contribution to $\chi$~\cite{Cole-Cole_41}.}.  
Both real and imaginary part
of $\chi(\omega, \,T)$ display a maximum for $\omega\tau\!=\!1$.
The basic experimental characterization of SCMs essentially reduces to determining the temperature dependence of 
$\xi$ and $\tau$, which is -- in principle -- possible thanks to equations~\eqref{xi_chi_stat} and~\eqref{chi_ac}. 

Even within the idealized scenario presented in this section, the way in which $\Delta_\xi$ and $\Delta_\tau$ defined in~\eqref{activation_xi-tau} 
relate to the Hamiltonian parameters $J$ and $D$ depends on the DW thickness, $w$, and on the temperature range in which such energy barriers are measured.    
Besides this, model Hamiltonians of real SCMs may differ significantly from~\eqref{Heisenberg_chain_classic}. 
In the next section we will recall some synthesis strategies that have been followed to produce different SCMs. 
The features of the employed building blocks and the type of coupling among them eventually decide which model is more appropriate to describe a specific SCM.

\section{Tailoring SCMs by building-block approach\label{mattoni}}   
The initial interest on Molecular Magnets stemmed from the attempt to design molecular systems displaying long-range magnetic order. However, after more than 30 years of attempts there are only two room-temperature molecular magnets and matters are no better for liquid-nitrogen temperatures~\cite{Miller-Gatteschi_CSR_2011}. 
To have long-range order it is necessary to build 2D or 3D structures of centres magnetically coupled. 
This is difficult with molecular bricks since the number of coordination sites which are available to propagate the exchange coupling in different directions is small due to the presence of capping ligands. Such bricks are then more suitable to produce low-dimensional systems, like clusters of metal ions (zero dimensional)~\cite{G-S-V} or spin chains. These systems do not display long-range order but still show a variety of interesting phenomena, including SCM behaviour.   
It is pedagogically useful to imagine that synthesizing a SCM is like assembling bricks with a magnetic functionality and a structural functionality. Usually, the latter is provided by organic molecules and the former by metal ions. Building blocks need to be chosen and arranged in a structure which maximizes the intrachain and minimizes the interchain interactions. Bricks with magnetic functionality must be coupled ferro- or ferri-magnetically and control of the magnetic anisotropy must be achieved.  
Chemists are not yet able to have that detailed control but progress is fast and serendipity always helps.

\begin{figure}
\includegraphics[width=11.cm]{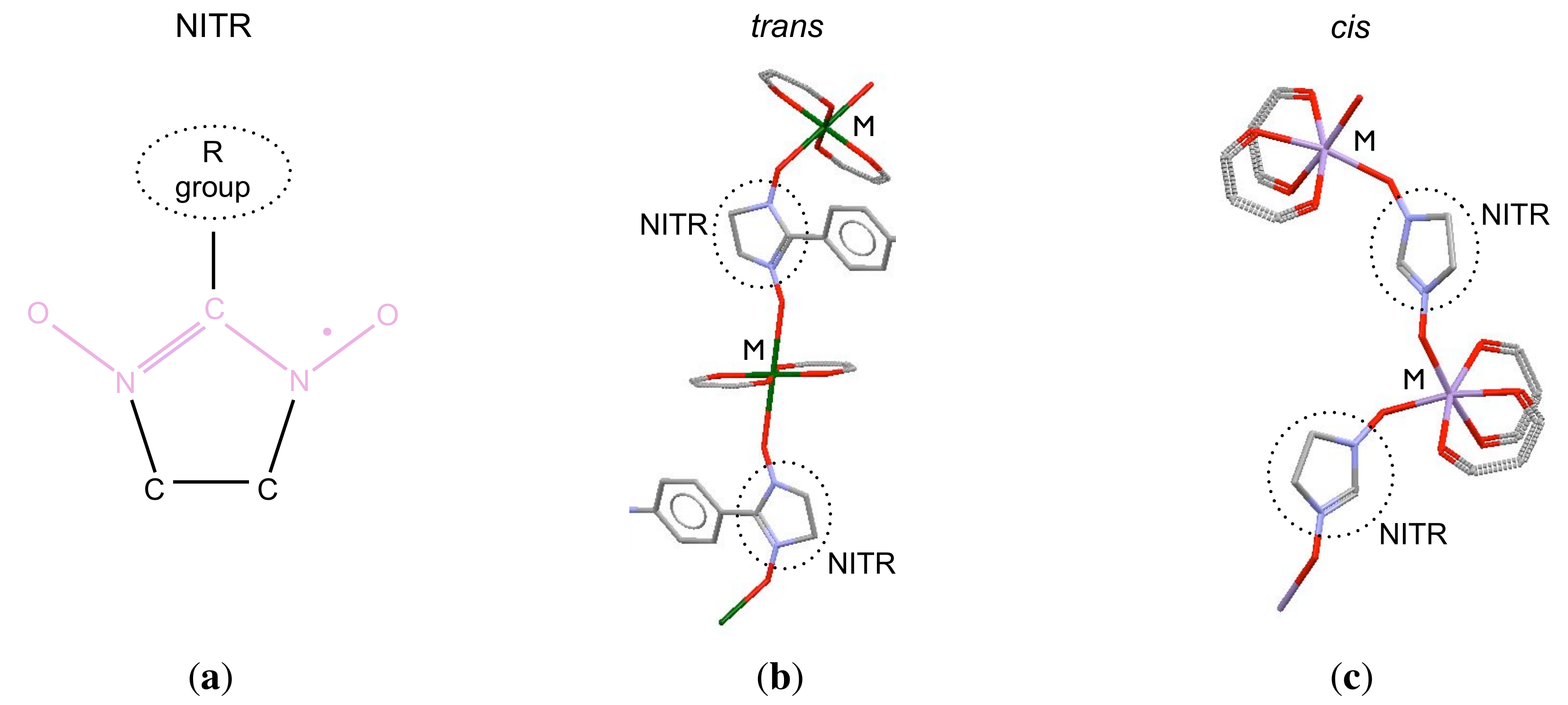} 
\vspace{-0.2cm}
\caption{ (\textbf{a}) schematic structure of the NITR radical: the unpaired electron is delocalised over the O-N-C-N-O fragment (coloured), which is magnetically active. 
(\textbf{b}) and (\textbf{c}) show two possible realisations of direct exchange coupling between the electron of each NITR radical and a metal ion (intrachain interaction):
each M can be bond to two NITR groups through oxygens occupying either \textit{trans} (\textbf{b}) or \textit{cis}  (\textbf{c}) positions in the 
coordination octahedron (see Fig.~\ref{metal-oxide}). }
\label{NITR_rad}
\end{figure}

Some centres of the building blocks shall be magnetically active, which implies the presence of unpaired electrons 
that are formally assigned to magnetic orbitals, either p, d, or f.  
In organic radicals the unpaired electrons normally belong to p orbitals: these are external orbitals which strongly interact with the environment. 
For this reason such electrons hardly remain unpaired but rather tend to couple with electrons of other molecules in covalent bonds,  
which eventually explains why few stable organic radicals exist.  
In the following, we will mostly refer to nitronyl nitroxide radicals (NITR), whose structure is shown in Fig.~\ref{NITR_rad}a. 
The unpaired electron is delocalised on the group O-N-C-N-O and, from the magnetic point of view, basically behaves as a free electron.  
Its magnetic moment is essentially spin determined, with little orbital contribution due to small spin-orbit coupling. This implies low magnetic anisotropy which is the final blow for purely organic SCMs. 

NITR radicals have the right geometry for bridging two metal ions through their equivalent oxygen atoms (extremes of the O-N-C-N-O fragment in Fig.~\ref{NITR_rad}a). 
The above considerations suggest that NITR radicals are not appropriate for being used alone, but they become excellent bricks for SCMs  
when coordinated to metal ions~\cite{Review_CC,Verdaguer_review,Review_Gao,JMC_Bogani,Yamashita_RSC_2012}. 
In fact, the interaction of the p orbitals with the d (or f) orbitals can be strong, of direct type, both ferro and antiferromagnetic in nature. 

\begin{figure}[ht]
\begin{center}
\includegraphics[width=11.5cm]{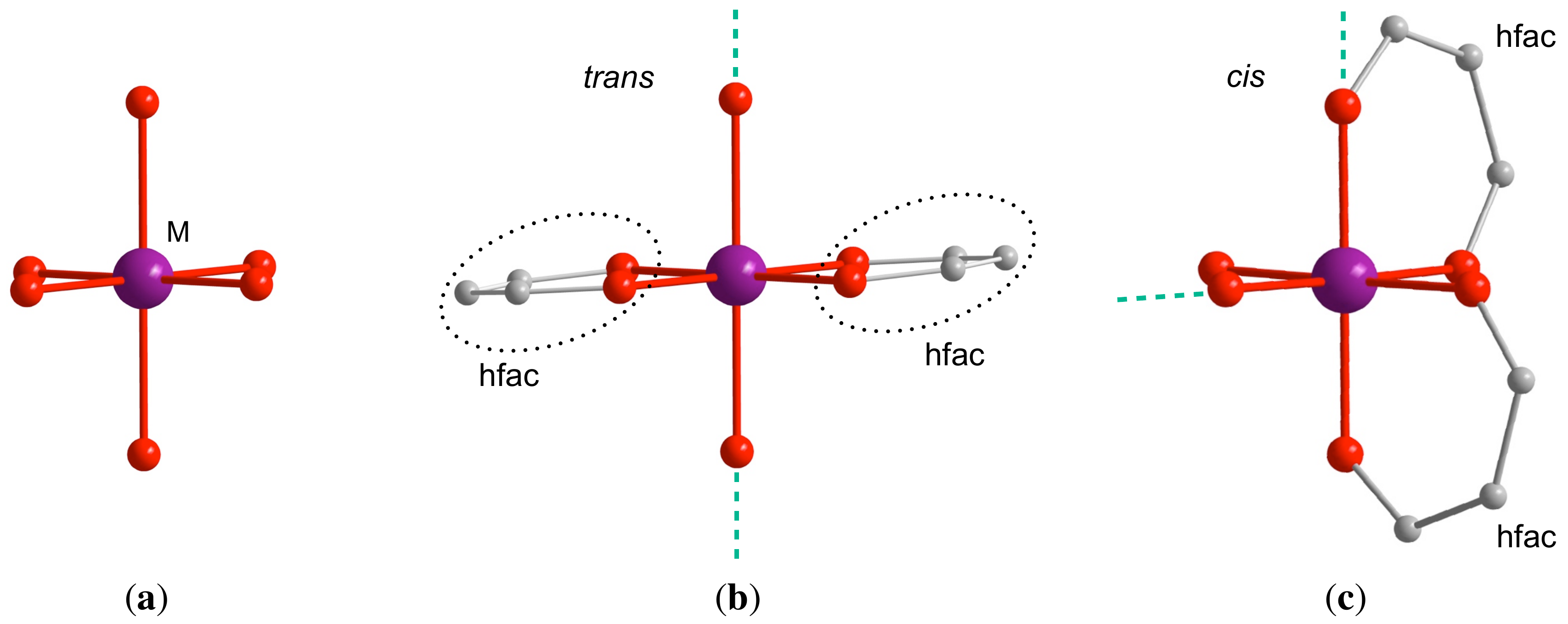}
\vspace{-0.2cm}
\caption{Sketch of a metal ion (purple spheres) in an octahedral environment of oxygen-donating ligands (red spheres representing oxygens).  (\textbf{a}) metal-oxide coordination in an extended solid. 
(\textbf{b}) and (\textbf{c}) M(hfac)$_2$ moiety with two empty coordination sites in \textit{trans} position (\textbf{b}) and in \textit{cis} position  (\textbf{c}); 
the CF3 groups of hfac ligands are not shown for clarity sake. Green dashed lines indicate the directions along which the intrachain exchange coupling mediated by a different ligand (e.g., NITR radical)
may propagate.  
\label{metal-oxide} }
\end{center}
\end{figure}

Transition-metal ions provide good magnetic bricks. As anticipated in the introduction, we will limit ourselves to consider SCMs possessing uniaxial anisotropy at the brick level. 
In molecular systems, magnetic anisotropy is closely related to the fact that the surrounding of metal ions is not spherically symmetric. 
Figure~\ref{metal-oxide}a shows a generic metal atom (M) in an octahedral environment of ligands. Oxygen atoms occupy the vertices of the octahedron. 
In the group M(hfac)$_2$, for instance, two oxygens of each hexafluoroacetylacetonate (hfac) ligand coordinate to M,  
thus occupying two neighbouring vertices of the octahedron per hfac molecule. 
The two remaining, empty coordination sites can be in either \textit{trans} or \textit{cis} position  
(Fig.~\ref{metal-oxide} b and c, respectively) and may host oxygens of other ligands that can be used to connect different M(hfac)$_2$ moieties.    
The choice of NITR to bridge those moieties creates a strong, direct exchange coupling between M and the electron delocalised on each O-N-C-N-O group (intrachain interaction, $J$). 
Consistently with the two possible coordination configurations of M(hfac)$_2$ sketched in Fig.~\ref{metal-oxide}, 
the segments connecting the metal ions with NITR oxygens may form an angle of 180$^\circ$ (\textit{trans}) or 90$^\circ$ (\textit{cis}).  
Such segments specify the direction along which the intrachain interaction propagates. 
The bulky hfac groups prevent efficient interchain exchange coupling.  
The residual interchain interaction $J'$ has mainly dipolar origin  and it is, typically, from 3 to 6 orders of magnitude smaller than the intrachain interaction. 
For this reason,  M(hfac)$_2$ moieties are perfectly suited for realizing isolated spin chains ($|J'/J| < 10^{-3}$).  
The interaction between successive magnetic bricks can either be ferro- or ferrimagnetic and give rise to straight or zig-zag structures. Besides, magnetic bricks are often characterized by low symmetry with the metal ions occupying general positions in the unit cells, which does not impose limitations to the orientation of anisotropy axes. Therefore, in practice, full collinearity among anisotropy axes is more an exception rather than the rule~\cite{Review_Gao,Review_CC,Verdaguer_review,Yamashita_RSC_2012}. 

Since SCM behaviour requires some magnetic anisotropy, the orbital momentum must not be completely quenched. The surviving component may 
be associated with single-ion anisotropy or with pair-spin interaction. In the former case, the residual orbital contribution can show up in a $\tens{g}$ tensor different from the free-electron one and/or in the zero-field splitting. With a large periodic table it is amazing that only cobalt and manganese, with some iron and nickel have been used. Mn$^{3+}$ is an example of anisotropy determined by zero-field splitting;  
while in Co$^{2+}$ the anisotropy is associated with the $\tens{g}$ tensor~\cite{CoII-JACS_08}. 
The crystal-field theory is the simplest way to describe the ground and low-lying levels of a transition-metal ion.  The Hamiltonian can be expressed as a sum of terms:
\begin{equation}
\label{Ham_Crystal-Field}
\mathcal{H} = \mathcal{H}_{0}+\mathcal{H}_\text{ee}+\mathcal{H}_\text{CF}+\mathcal{H}_\text{LS} \;,
\end{equation}
where $\mathcal{H}_{0}$ is the origin of the electron configuration (3d)$^n$, $\mathcal{H}_\text{ee}$ is the electron-electron repulsion, 
$\mathcal{H}_\text{CF}$ is the crystal-field term and $\mathcal{H}_\text{LS}$ is the spin-orbit coupling. 
For 3d ions $\mathcal{H}_\text{ee}$ and $\mathcal{H}_\text{CF}$ are comparable and larger than spin-orbit coupling. It is customary to neglect in first approximation the spin-orbit coupling which is introduced later as a perturbation.
Mn$^{3+}$ has a (3d)$^4$ valence-electron configuration which in octahedral symmetry yields a $^5$E$_\text{g}$ ground state. This is unstable to phonon coupling (Jahn-Teller theorem), 
which lowers the symmetry to D$_\text{4h}$, namely to a tetragonally elongated coordination. The ground state $^5$A$_\text{1g}$, in zero-order approximation, is five-fold degenerate
(no orbital degeneracy and $S$=2 spin multiplet). The spin-orbit \textit{perturbation} yields no contribution in the first order, but to the second order it admixes excited states with the ground state. 
This removes the degeneracy of the spin multiplet and produces anisotropic components in the $\tens{g}$ tensor. 
Its effect is usually summarized introducing an effective single-ion spin Hamiltonian of the form:   
\begin{equation}
\label{Spin_Ham}
\mathcal{H} = - D\hat{S}_z^2 - \mu_\text{B} \vec{B} \tens{g}  \hat{\vec S}	\;,
\end{equation}
where $\tens{g}$  is a symmetric tensor.  
The first term is responsible for the zero-field splitting of the 2$S$+1 levels. It is often referred to as crystal-field term, even though this is misleading because it is not the crystal field which splits the levels but rather the spin-orbit coupling. 
The spin-Hamiltonian parameters are determined by the spin-orbit coupling constant $\lambda$ and by the degree of mixing between the $^5$A$_\text{1g}$ ground state and the excited states 
induced by $\mathcal{H}_\text{LS}$~\cite{Barra_97}.     
The lowering of the symmetry produces axially symmetric $\tens{g}$ and $\tens{D}$ tensors%
\footnote{Without loss of generality $\tens{g}$ and $\tens{D}$ can be assumed symmetric. 
Consequently, they are diagonal on a proper reference frame with eigenvalues $g_{x}$, $g_{y}$, $g_{z}$ and $D_{x}$, $D_{y}$, $D_{z}$. 
The same notation will be used for the $\tens{J}$ tensor.}  
whose components are related, to the leading order, through the following formula: 
\begin{equation}
\label{D-g_relation}
D=D_{z}-D_{x,y}=\lambda\left(g_{x,y}- g_{z}\right)\;.
\end{equation}

\section{Realistic spin Hamiltonians for Single-Chain Magnets\label{Hamiltonian}}
So far we have neglected the coupling among spin pairs, which can be written as
\begin{equation}
\label{J_jk}
\mathcal{H}_\text{exch}= -\hat{\vec S}_{p}\tens{J}\hat{\vec S}_{k}\;,
\end{equation}
where  $\hat{\vec S}_{p}$ and  $\hat{\vec S}_{k}$ are effective spin operators of any two interacting magnetic bricks and $\tens{J}$ is a generic 3-by-3 matrix. 
Limiting ourself to intrachain spin-spin coupling, we can neglect the contribution due to dipolar interaction which is typically much smaller than the exchange one. 
When pair-spin interaction involves transition metals whose ground state is not orbitally degenerate, the isotropic contribution to the $\tens{J}$ tensor 
dominates. As mentioned before, second-order perturbation theory prescribes 
that the ground-state wave functions be modified because of the admixing with excited states mediated by spin-orbit coupling.   
When the corrected wave functions of the bricks $p$ and $k$ are employed 
to compute the exchange integral, the anisotropic and antisymmetric contributions emerge. The former is proportional to $\left(\Delta g/g\right)^2$, while the latter
is proportional to $\Delta g/g $. This ratio is usually much smaller than one, thus the antisymmetric term -- if allowed by symmetry -- 
is expected to dominate with respect to the anisotropic exchange. For our purposes, it will be enough to know that anisotropic contributions to $\tens{J}$ can be 
neglected when the $\tens{g}$ tensor is fairly isotropic, as for Mn$^{2+}$, Mn$^{3+}$, high-spin Fe$^{3+}$, etc.  

When magnetic bricks comprise transition metals with orbitally degenerate ground state, predicting the properties of the $\tens{g}$, $\tens{D}$ and $\tens{J}$ tensors  
on simple footing becomes extremely complicated~\cite{Modesto-Dunbar-Coronado_CSR_2011}. 
One possibility is that of considering just a symmetric exchange tensor, which is then diagonal on a proper basis with principal values 
$J_x$, $J_y$ and $J_z$. If compatible with symmetry, an antisymmetric, Dzyaloshinskii-Moriya term may be added independently.  

In 1D magnetic systems realized by coupling radicals with neighbouring transition-metal ions, anisotropic terms in $\tens{J}$ may originate only from the metal atoms.     
The first successful examples consisted in ferrimagnetic chains of general formula Mn(hfac)$_2$NITR~\cite{Kai_IC_89}. The radical is isotropic and so is Mn$^{2+}$, 
therefore $\tens{J}$ is expected to be proportional to the identity. Indeed, these systems represented text-book examples of 1D Heisenberg ferrimagnets 
described by the Hamiltonian  
\begin{equation}
\label{Seiden}
\mathcal{H}_\text{Mn-rad}= -J\sum_{ p=1}^{N/2}\hat{\vec S}_{2p}\cdot\hat{\vec s}_{2p+1}\;,
\end{equation}
where $\hat{\vec S}_{2p}$ stand for Mn$^{2+}$ spin operators (lying at even sites $2p$ with $S_{2p}=5/2$), while $\hat{\vec s}_{2p+1}$ are the radical spin-one-half operators. $J$ is  negative and tends to orient the  nearest-neighbouring spins antiparallel to each other. The temperature dependence of the static susceptibility was fitted using the Seiden model~\cite{Seiden} with $|J|$ in the range 300--475 K depending on the 
substituent R on the radical%
\footnote{Henceforth, energies will be expressed in Kelvin units to make it easier to compare them with thermal energy. 
The conversion factor to SI coincides with the Boltzmann constant $k$: 1 K = 1.3806503$\times$10$^{-23}$ J.}~\cite{Kai_IC_88}. 
In the Seiden model the Mn spins are replaced by classical vectors, which -- in the absence of field and single-ion anisotropy -- makes the model analytically solvable.  
Due to the large value of the coupling between Mn$^{2+}$ and NITR radicals, strong pair-spin correlations develop, which is highlighted by a divergence of the  correlation length at low temperature (proportional to $|J|/T$ and not exponential like in spin chains with uniaxial anisotropy~\cite{Kai_IC_88}).  
In the presence of such strong intrachain correlations even a tiny interchain interaction $J'$ may induce 3D ordering~\cite{Scalapino_quasi_1d}.  
In Mn(hfac)$_2$(NITiPr) this happens at $T_\text{C}$=7.6 K~\cite{Kai_IC_89}.   
ESR and NMR studies provided evidence of spin-diffusion effect allowing for an estimate of the ratio between inter- and intrachain exchange interaction of the order $|J'/J|$=2$\times$10$^{-6}$~\cite{Ferrero_JACS_91}.    \\
This example confirms that combining transition metals with organic radicals is a powerful strategy for designing ideal 1D systems. 
An additional ingredient is needed to realize a SCM: magnetic anisotropy. This may easily be introduced by replacing Mn$^{2+}$ with Mn$^{3+}$. 
Recently, the observation of slow relaxation consistent with SCM features was reported for ferrimagnetic spin chains consisting of  
Mn$^{3+}$ and TCNQ or TCNE\footnote{Acronyms stand for tetracyanoquinodimethane (TCNQ) and tetracyanoethylene (TCNE).} 
organic radicals~\cite{Large-DW_SCMs,Yamashita_RSC_2012}. The relatively large multiplicity of Mn$^{3+}$ spins, $S$=2, 
allows justifying their replacement by classical vectors. Thus, the Seiden model is still a good starting point for describing the 
magnetic properties of these systems, provided that single-ion-anisotropy terms are added.  
Even if the modelling aspects are well-defined, the rationalization of Mn$^{3+}$-radical SCMs is complicated by the fact that $J \gg D$, meaning that the relevant excitations are broad DWs. 

The extreme anisotropic $\tens{g}$ tensor obtained for Co$^{2+}$ in a tetragonally compressed symmetry suggests that its coupling with NITR be, to leading order, of the Ising type.  
This idea led to the synthesis of the first compound showing SCM behaviour: Co(hfac)$_2$(NITPhOMe)~\cite{Angew}.    
Experimental results pertaining slow dynamics have shown a substantial agreement with the kinetic version of the Ising model developed by Glauber~\cite{Glauber}.    
Unfortunately, up to date, the static properties have not been successfully modelled yet. The first reason is that above 40 K 
treating Co$^{2+}$ as an effective $S=$1/2 is not legitimate (the energy separation between the ground-state Kramers doublet and the excited multiplets is about 100 K). 
A second reason relates to the helical structure of this compound, because of which the elementary magnetic cell contains 3 Co$^{2+}$ and 3 radical spins. 
Apart from the question of reproducing its static properties, it is instructive to give a closer look at the Hamiltonian of this system to show how 
non-collinearity can be modelled in general.  
For temperatures lower than 40 K, a reasonable Hamiltonian for the Co(hfac)$_2$(NITPhOMe) chain is given by 
\begin{equation}
\label{Co-rad}
\mathcal{H}_\text{Co-rad}= 
-\sum_{p=1}^{N/N_r}\sum_{r=1}^{N_r/2}\left[\hat{\vec S}_{p,2r}\tens{J}_{2r}\hat{\vec s}_{p,2r+1} 
+\mu_\text{B}\vec B\tens{g}_{2r}\hat{\vec S}_{p,2r}
+g\vec B\cdot\hat{\vec s}_{p,2r+1} \right]\;,
\end{equation} 
where both $\hat{\vec S}_{p,2r}$ and $\hat{\vec s}_{p,2r+1}$ are spin one-half operators associated with Co$^{2+}$ ions and radicals, respectively.  
$p$ represents the magnetic cell index while $r$ spans the inequivalent Co$^{2+}$ atoms inside each cell. For the specific case, $r$ takes $N_r$=3 different values 
which correspond to different orientations of the principal axes along which the $\tens{J}$ and $\tens{g}$ tensors are diagonal. 
If spin projections are expressed in the crystal frame, the tensors appearing in Hamiltonian~\eqref{Co-rad} are built applying a standard $O(3)$ rotation to the diagonal tensors~\cite{PRB_Luzon,Eur_Phys_Lett,CoII-JACS_08}.  
Formally, $r$ in $\tens{J}_{2r}$ and $\tens{g}_{2r}$ labels different sets of rotation angles.     
The Land\'e factor of the radical is isotropic and thus independent of $r$.  \\
When spins  $S>1/2$ are considered, a magnetic brick may possess some single-ion anisotropy, which implies that also the $\tens{D}$ tensor needs to be rotated in non-collinear systems.

The thermodynamic properties of classical spin chains with nearest-neighbour interactions  
can be efficiently computed by means of the transfer-matrix method. Letting the general Hamiltonian be 
$\mathcal{H}=-kT\sum_{p}V(\vec{S}_{p},\, \vec{S}_{p+1})$, the partition function $\mathcal{Z}$ is 
obtained integrating over all the possible directions along which each unitary vector $\vec{S}_{p}$ may point: 
\begin{equation}
\label{zeta_TM_general}
\mathcal{Z}=\int d\Omega_{1}\int
d\Omega_{2}\ldots \int \E^{V(\vec{S}_{1},\, \vec{S}_{2})}\,
\E^{V(\vec{S}_{2},\, \vec{S}_{3})}\, \ldots \, \E^{V(\vec{S}_{N},\, \vec{S}_{1})}\, d\Omega_{N}\;.
\end{equation}
Defining the transfer kernel as $\mathcal{K}(\vec{S}_{p},\, \vec{S}_{p+1})=\E^{V(\vec{S}_{p},\, \vec{S}_{p+1})}$
and assuming periodic boundary conditions, the partition function $\mathcal{Z}$ can be recasted into the trace of the $N$-th power
of $\mathcal{K}(\vec{S}_{p},\, \vec{S}_{p+1})$:
\begin{equation}
\label{TM_kernel_trace}
\mathcal{Z}=\int d\Omega_{1}\int
d\Omega_{2}\ldots \int \mathcal{K} (\vec{S}_{1},\, \vec{S}_{2})\,
\mathcal{K} (\vec{S}_{2},\, \vec{S}_{3})\, \ldots \,
\mathcal{K}(\vec{S}_{N},\, \vec{S}_{1})\, d\Omega_{N}= \rm Tr\big\{
\mathcal{K}^{N}\big\} \;.
\end{equation}
When the transfer kernel is expressed on a basis of eigenfunctions, the partition function reduces to a sum of eigenvalues
$ \mathcal{Z}=\sum_{m}\lambda_{m}^{N}$, where $\psi_{m}(\vec{S}_{p})$ and $\lambda_{m}$ are solutions of the following eigenvalue problem: 
\begin{equation}
\label{TM_integral equation_dex} 
\int \mathcal{K}(\vec{S}_{p},\,
\vec{S}_{p+1})\psi_{m}(\vec{S}_{p+1}) d\Omega_{p+1}=\lambda_{m}
\psi_{m}(\vec{S}_{p})\;.
\end{equation}
For kernels that can be written in a symmetric form with respect to the exchange  $\vec{S}_{p} \leftrightarrow \vec{S}_{p+1}$ 
the spectral theorem warrants that eigenvalues are real%
\footnote{In the general, non-symmetric case left and right eigenfunctions of $\mathcal{K}(\vec{S}_{p},\, \vec{S}_{p+1})$ 
have to be considered, but the basic ideas of the transfer-matrix method remain the same.}.  
They are also  positive, because the transfer kernel is a positive function of $\vec{S}_{p}$ and $\vec{S}_{p+1}$, and upper bounded so that they can be ordered from the largest to the smallest one: $ \lambda_{0}>\lambda_{1}>\lambda_{2}>\ldots$ In the thermodynamic  limit the asymptotic behaviour of the
partition function~\eqref{TM_kernel_trace} is dominated by the
largest eigenvalue $\lambda_{0}$, $\mathcal{Z}\simeq \lambda_{0}^{N}$,  meaning that the free energy per spin is given by  $f=-kT \log \lambda_0$.  
Macroscopic experimental observables are obtained as derivatives of $f$, but this method allows computing 
microscopic averages as well. Apart from some fortunate cases~\cite{Fisher}, equation~\eqref{TM_integral equation_dex} needs to be solved numerically 
by sampling the unitary sphere with a finite number of special points. This number can be increased dynamically untill the desired precision is reached.  
Even though it may not be transparent from our description, a new eigenvalue problem ought to be solved for any computed temperature or applied field.   
Referring the reader to the existing literature for implementation details~\cite{APA_Vindigni,Blume-Tannous,PRB_Luzon}, we remark that the transfer-matrix method allows computing the 
magnetic properties of classical spin chains more efficiently than, e.g., standard Metropolis MonteCarlo. 
This makes it possible to fit spin Hamiltonian parameters directly to 
experimental data sets. The major drawback is that the number of spin variables that appear in the kernel scales like the range of interaction 
(2 for nearest-neighbour, 4 for second nearest-neighbour interaction, etc.), which finally affects the complexity of the eigenvalue problem in~\eqref{TM_integral equation_dex}. 

The transfer-matrix method can easily be extended to models in which classical and quantum spins alternate, like in the Seiden model~\cite{Seiden}.  
Noting that the quantum-spin operators are not directly coupled with each other, one can integrate out their degrees of freedom independently. 
In fact, a generic quantum spin located at site $2p+1$ experiences an effective ``field''
$k T\vec h_{2p,2p+2}= J\left(\vec S_{2p}+\vec S_{2p+2}\right)+\mu_\text{B}g\vec B$. The corresponding energy levels are $\pm k T |\vec h_{2p,2p+2}|$, 
which depend \textit{parametrically} on the orientation of the two classical spins, $\vec S_{2p}$ and $\vec S_{2p+2}$.  
After tracing over the quantum degrees of freedom, one is left with the kernel  
\begin{equation}
\label{Mn$^{3+}$_kernel}
\mathcal{K}(\vec{S}_{2p},\, \vec{S}_{2p+2})=2\cosh\left(|\vec h_{2p,2p+2}|\right)\exp\left(\frac{\mu_\text{B}\vec B\tens{g}\vec S_{2p}}{k T}\right)
\exp\left(\frac{D(S_{2p}^z)^2}{k T}\right)
\end{equation}
where the single-ion anisotropy and Zeeman term acting on the classical spins have been added. 
The kernel~\eqref{Mn$^{3+}$_kernel} may be used to compute, e.g., the equilibrium suceptibility of Mn$^{3+}$-radical chains~\cite{Large-DW_SCMs}. 
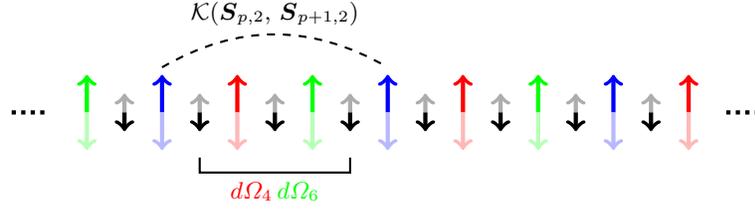
\begin{figure}
\vspace{-0.4cm}
\begin{center}
\begin{tikzpicture}[scale=1, ultra thick]
	\foreach \t in {0,3,6}
	\foreach \x/\c in {0/green,1/blue,2/red}
	\draw [->,\c] ({\t+\x},0.) -- ({\t+\x},0.5);
	\foreach \t in {0,3,6}
	\foreach \x/\c in {0/green!28!white,1/blue!28!white,2/red!28!white}
	\draw [<-,\c] ({\t+\x},-0.5) -- ({\t+\x},0.);
	\foreach \p in {0.5,1.5,...,7.5}
	\draw [->,black!32!white] (\p,0) -- (\p,0.25);
	\foreach \p in {0.5,1.5,...,7.5}
	\draw [<-,black] (\p,-0.25) -- (\p,0);
	\draw[dotted] (-1,0) -- (-0.5,0) (8.5,0) -- (9,0);
	\draw [thick](1.5,-0.6) -- (1.5,-0.8) -- (3.5,-0.8)  node [below, midway] {$\textcolor{red}{d\Omega_4}\, \textcolor{green}{d\Omega_6}$} -- (3.5,-0.6);
	\draw [dashed, thick, bend left] (1,0.6) to  node [above, midway] {$\mathcal{K}(\vec{S}_{p,2},\, \vec{S}_{p+1,2})$} (4,0.6);
\end{tikzpicture}
\end{center}
\caption{Sketch of the periodicity associated with Hamiltonian~\eqref{Co-rad}: small, black arrows represent radical spins $\hat{\vec s}_{p,2r+1}$, while the large coloured arrows represent the metal-ion ones $\hat{\vec S}_{p,2r}$. The kernel~\eqref{Three-fold_kernel} depends only on  
$\vec{S}_{p,2}$ and $\vec{S}_{p+1,2}$ (blue arrows) because an integration over the internal dgrees of freedom $\vec{S}_{p,4}$  (red arrows) 
and $\vec{S}_{p,6}$  (green arrows) has been performed. \label{Three-fold_fig}} 
\end{figure}
To the aim of sketching how to proceed for modelling non-collinearity, let us substitute the spins $\hat{\vec S}_{p,2r}$ 
in Hamiltonian~\eqref{Co-rad} by classical vectors
\footnote{Even though this is not justified for the specific case of Co$^{2+}$, the classical approximation allows us to discuss the general formalism.}. 
Even after integrating out the radical degrees of freedom, the are still 3 \textit{non-equivalent} classical spins in each magnetic unit cell, 
resulting in 3 different kernels if $\vec B$ is applied along a generic direction: $\mathcal{K}(\vec{S}_{p,2},\, \vec{S}_{p,4})$,   
$\mathcal{K}(\vec{S}_{p,4},\, \vec{S}_{p,6})$ and $\mathcal{K}(\vec{S}_{p,6},\, \vec{S}_{p+1,2})$. The role of the kernel~\eqref{Mn$^{3+}$_kernel} 
is played by  
\begin{equation}
\label{Three-fold_kernel}
\mathcal{K}(\vec{S}_{p,2},\, \vec{S}_{p+1,2})=
\int  d\Omega_{4}\int\mathcal{K}(\vec{S}_{p,2},\,\vec{S}_{p,4})\,
\mathcal{K}(\vec{S}_{p,4},\,\vec{S}_{p,6})\,\mathcal{K}(\vec{S}_{p,6},\,\vec{S}_{p+1,2})\,d\Omega_{6}\;,
\end{equation}
obtained by tracing over the degrees of freedom internal to the considered cell, $d\Omega_{4}$ and $d\Omega_{6}$%
\footnote{Actually, the choice of the unit cell is not unique: one might integrate over any 
pair of internal degrees of freedom $d\Omega_{2r}$. This turns necessary in order to compute microscopic averages of 
individual spin components.}.  
The way in which the kernel is built is sketched pictorially in Fig.~\ref{Three-fold_fig}.  
In the thermodynamics limit the partition function is given by $\mathcal{Z}\simeq \lambda_{0}^{N/N_r}$, where the number of spins have been replaced by the number of unit cells $N/N_r$. \\
Due to non-collinearity, a strong anisotropy in the $\tens{D}$, $\tens{g}$ or $\tens{J}$ tensors may not necessarily be evident at the macroscopic level~\cite{Eur_Phys_Lett}. 
More concretely, having similar saturation values for the magnetisation along different crystallographic directions may still be compatible with a strong uniaxial character at the level
of individual bricks. Non-collinearity is also consistent with an inversion of the directions of easy and hard magnetisation by increasing temperature~\cite{PRB_Luzon} or with the vanishing of the correlation length 
for some specific applied fields~\cite{PRB_Vindigni_04}.   \\
In passing, we note that finite-size effects can be taken into account in the general transfer-matrix framework~\cite{APA_Vindigni} 
as well as interchain interactions if treated at the mean-field level~\cite{Ruffo-Mukamel,Scalapino_quasi_1d}.   

\begin{figure}[ht]
\vspace{-0.6cm}
\begin{center}
\begin{tikzpicture}[scale=1, ultra thick]
	\draw [fill=green!20!white,thin](0.5,0)--(1,0) arc (0:36:0.5);
	\draw [thin] (1.05,0.22) node [right,scale=1.1] {$\theta$};
	\draw [fill=green!20!white,thin](2.5,0)--(3,0) arc (0:-36:0.5);
	\draw [thin] (3.05,-0.22) node [right,scale=1.1] {$\theta$};
	\foreach \p in {0.5,4.5,...,8.5}
	\draw [->,blue] (\p,0) -- +(36:1);
	\foreach \p in {0.5,4.5,...,8.5}
	\draw [->,blue!28!white] (\p,0) -- +(216:1);
	\foreach \p in {2.5,6.5,...,10.5}
	\draw [->,red] (\p,0) -- +(-36:1);
	\foreach \p in {2.5,6.5,...,10.5}
	\draw [->,red!28!white] (\p,0) -- +(144:1);
	\draw[->,thick] (-0.2,0) -- (11.5,0);
	\draw [thin] (11.2,0)  node [above,scale=1.3] {$\underline c$};
	\draw [thin] (4.8,0.8) node [scale=1.3] {$z_\text{2r}$};
	\draw [thin] (6.7,-0.8) node [scale=1.3] {$z_\text{2r+1}$};
	\foreach \ba in {-1.15}
	\draw [thin,white](2+\ba,-2.)--(3.3+\ba,-2.)node[above,midway,scale=1.3]{\textcolor{green}{\boldmath{$B$}$_\parallel$}};
	\foreach \ba in {-1.15}
	\draw [thin,white](2+\ba,-2.3)--(3.3+\ba,-2.3)node [below,midway,scale=1.1]{\textcolor{green}{slow relaxation}};
	\foreach \ba in {-1.15}
	\draw [thin,fill=green] (2+\ba,-2.)--(3+\ba,-2.)--(3+\ba,-2.2)--(2+\ba,-2.2)--(2+\ba,-2.) (3+\ba,-1.9)--(3+\ba,-2.3)--(3.3+\ba,-2.1)--(3+\ba,-1.9);
	\foreach \ba in {0.4}
	\draw [thin,white](9.3+\ba,-2.2)--(9.3+\ba,-0.9)node[right,midway,scale=1.3]{\textcolor{gray}{\boldmath{$B$}$_\perp$}};
	\foreach \ba in {0.4}
	\draw [thin,white](9+\ba,-2.3)--(9.2+\ba,-2.3)node [below,midway,scale=1.1]{\textcolor{gray}{fast relaxation}};
	\foreach \ba in {0.4}
	\draw [thin,fill=gray](9+\ba,-2.2)--(9.2+\ba,-2.2)--(9.2+\ba,-1.2)--(9+\ba,-1.2)--(9+\ba,-2.2) (8.9+\ba,-1.2)--(9.3+\ba,-1.2)--(9.1+\ba,-0.9)--(8.9+\ba,-1.2);
\end{tikzpicture}
\end{center}
\caption{Sketch of a two-fold non-collinear Ising chain. The local anisotropy axes have been chosen coplanar for simplicity and form an angle $\theta$ with the chain axis 
$\underline c$. For $\theta\!<\!\pi/4$ and $J\!>\!0$, slow relaxation is expected only when $\vec B$ is applied parallel to the chain axis. 
\label{two-fold-chain}}
\end{figure}
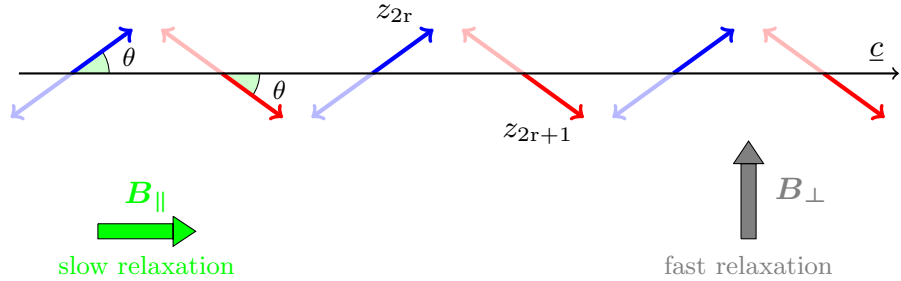

In the cases in which one of the principal values of the  $\tens{D}$ or $\tens{J}$ tensors is much larger than the other two (say $J_z\gg J_x,J_y$), 
spin operators can be substituted by two-valued classical variables $\sigma_{p} = \pm 1 $. In this way, the problem reduces to the Ising Hamiltonian  
\begin{equation}
\label{Ising_Hamiltonian_in_Glauber} 
\mathcal{H}=-J\sum ^{N}_{p=1} \, \sigma_{p}\sigma_{p+1} 
-\mu_\text{B} B\sum ^{N}_{p=1}g_{p}\sigma_{p} \;,
\end{equation}
in which $J$ and $g_p$ may contain information about non-collinearity. In Fig.~\ref{two-fold-chain} a sketch of a two-fold, non-collinear Ising chain is shown. 
Assuming that both the $\tens{g}$ and $\tens{J}$ tensors have only one non-zero component along their principal axes, 
the corresponding parameters in Hamiltonian~\eqref{Ising_Hamiltonian_in_Glauber} are given by $J\!=\!\cos(2\theta)J_z$ and $g_p\!=\!\cos(\theta)g_z$ if $\vec B$ is parallel 
to the chain axis ($\underline{c}$ axis in Fig.~\ref{two-fold-chain}) while $g_p\!=\!(-1)^p\sin(\theta)g_z$ if $\vec B$ is perpendicular to the chain%
\footnote{The reader is addressed to Ref.~\cite{Modesto-Dunbar-Coronado_CSR_2011,CoII-JACS_08} for a more rigorous treatment.}.   
The framework in which static properties of SCMs can be modelled seems to be well-defined. However, it should not be forgotten that 
the genuine 1D static behaviour can be accessed only above a certain temperature $T_\text{b}$, dependent on the specific experiment, below which 
slow dynamics starts playing a major role. 
Moreover,  finite-size effects or 3D interchain interactions may come into play at higher temperature than $T_\text{b}$~\cite{Coulon_PRL_09}.   
On the high-temperature side, distinctive 1D features (short-range order) smear out in the isotropic paramagnetic phase.  
All these phenomena set limitations to the applicability of any equilibrium 1D model.

\section{Glauber model and Single-Chain Magnets\label{Glauber}}
In this section we will assume the anisotropy at the brick level to be large enough that Hamiltonian~\eqref{Ising_Hamiltonian_in_Glauber} 
suffices to discuss the important features of slow dynamics. 
A kinetic version of the Ising model was proposed by J. R. Glauber in 1963~\cite{Glauber}. As based on stochastic dynamics, this model refers to coarse-grained dynamics, typically 
some orders of magnitude longer than a Larmor period.   

Following Glauber, let $P(\underline{\sigma},t)$ be the probability 
of occurrence of some  configuration $\underline{\sigma}=\sigma _{1},...,\sigma_{p},...,\sigma _{N} $ at time $t$ 
and $w_{\sigma_{p}\rightarrow-\sigma_{p}}$ the probability of reversing the $p$-th spin per unit time. 
The master equation of the problem, thus, reads 
\begin{equation}
\label{Master_Equation_Glauber}
\begin{split}
\frac{\D}{\D t}P(\underline{\sigma},t)&=-\sum _{p=1}^{N}\, w_{\sigma_{p}\rightarrow -\sigma _{p}}P(\sigma_{1},.....,\sigma_{p},.....,\sigma _{N},t)+\\
&+\sum ^{N}_{p=1}\, w_{-\sigma_{p}\rightarrow \sigma _{p}}P(\sigma _{1},.....,-\sigma_{p},.....,\sigma _{N},t)\;.
\end{split}
\end{equation}
To model the magnetisation dynamics and a.c. susceptibility it is not necessary to solve equation~\eqref{Master_Equation_Glauber}: one can limit oneself to single-spin averages $s_p(t)$ 
\begin{equation}
s_{p}(t) =\sum_{\left\{ \underline{\sigma} \right\} }  \sigma_{p} P (\underline{\sigma},t ) \;.
\end{equation}
It can be shown that spin averages fulfil the differential equation 
\begin{equation} 
\label{Glauber_eq_Motion_1}
\frac{\D s_{p}}{\D t}=-2\left\langle\sigma_{p} w_{\sigma_{p}\rightarrow -\sigma _{p}}\right\rangle\;,
\end{equation}
where $\langle \dots \rangle$ denotes, again, the time-dependent average performed by means of $P(\underline{\sigma},t)$.
In order that dynamics drives the system towards Boltzmann equilibrium, the detailed-balance condition shall hold 
\begin{equation} 
\label{Glauber_DB}
\frac{P_\text{eq}(\sigma _{1},...,-\sigma
_{p},...,\sigma _{N})}{P_\text{eq}(\sigma _{1},...,\sigma _{p},...,\sigma
_{N})}=\frac{w_{\sigma _{p}\rightarrow -\sigma _{p}}}{w_{-\sigma
_{p}\rightarrow \sigma _{p}}}\;.
\end{equation} 
The equilibrium probabilities on the left-hand side of~\eqref{Glauber_DB} are obtained reversing the $p$-th spin while leaving the other ($N-1$) unchanged.  
Their ratio can be written as follows 
\begin{equation}
\begin{split}
&\frac{P_\text{eq}(\sigma _{1},...,-\sigma _{p},...,\sigma_{N})}{P_\text{eq}(\sigma _{1},...,\sigma _{p},...,\sigma
_{N})}=\frac{\exp \left[ -\kappa\sigma_{p}\left(\sigma_{p-1}+\sigma _{p+1}\right) \right] \exp \left( -h_p\sigma_{p}\right)}
{\exp \left[ \kappa \sigma_{p}\left( \sigma_{p-1}+\sigma_{p+1}\right) \right] \exp \left( h_k\sigma_{p}\right)}\\
&\quad= \frac{ \left[ 1-\frac{1}{2}\sigma_{p}\left(\sigma _{p-1}+\sigma_{p+1}\right) \tanh ( 2\kappa) \right]\left[1-\sigma_{p}\tanh (h_p) \right]}
{\left[ 1+\frac{1}{2}\sigma_{p}\left(\sigma_{p-1}+\sigma_{p+1}\right) \tanh ( 2\kappa ) \right]\left[1+ \sigma_{p}   \tanh (h_p) \right]  }
\end{split}
\end{equation}
with $\kappa=J/k T$ and $h_p=\mu_\text{B} B g_p /k T$. The above relation suggests the following form for the transition probability:  
\begin{equation}
\label{Transition_Prob_Field_Glauber} 
w_{\sigma_{p}\rightarrow -\sigma _{p}}=\frac{1}{2}\alpha \left[
1-\frac{1}{2}\gamma \sigma _{p}\left( \sigma _{p-1}+\sigma
_{p+1}\right) \right] \left[ 1-  \sigma _{p}\tanh(h_p) \right]\;,
\end{equation}
with $ \gamma =\tanh (2\kappa)$, so that detailed balance~\eqref{Glauber_DB} is automatically fulfilled. 
Equation~\eqref{Transition_Prob_Field_Glauber} corresponds to Glauber's original choice; other transition probabilities fulfilling~\eqref{Glauber_DB} could 
be chosen~\cite{Facilitated_dyn}, but -- to our knowledge -- they have not been considered in the context of SCMs. 
Note that the parameter $\alpha$ entering~\eqref{Transition_Prob_Field_Glauber} sets the natural time unit of the model. 
It can be interpreted as the attempt frequency of an isolated spin, i.e., the transition probability for vanishing exchange coupling, $J=0$. 
Already Suzuki and Kubo commented that, in general, $\alpha$ should depend on temperature~\cite{Suzuki-Kubo_67}. We will come back to this important point further on.
Combining~\eqref{Glauber_eq_Motion_1} and~\eqref{Transition_Prob_Field_Glauber}, a set of differential equations for spin averages is obtained
\begin{equation} 
\label{Glauber_eq_Motion_2}
\frac{1}{\alpha}\frac{\D s_{p}}{\D t}=-\left[s_{p}-\frac{\gamma}{2}\left(s_{p-1}+s_{p+1} \right)\right] 
+\left[1-\frac{\gamma}{2} \left(r_{p-1,p}+r_{p,p+1} \right)\right]\tanh(h_p)\;,
\end{equation}
where $r_{p,l}=\langle\sigma_p \sigma_l\rangle$. This means that the knowledge of pair-spin correlations is needed to solve~\eqref{Glauber_eq_Motion_2}.  
In turn, the knowledge of three-spin correlations is needed to obtain $r_{p,l}$ and so on. 
In other words, equation~\eqref{Glauber_eq_Motion_2} is the first one of an infinite hierarchy of kinetic equations~\cite{Glauber,Huang_PRA_73}. 
A judicious truncation of this series is, therefore, required in order to get analytic results which could easily be compared with experiments. 
In the following, we will analyse different decoupling schemes, related to different physically relevant situations. 
Another crucial point concerns the choice of boundary conditions for the system~\eqref{Glauber_eq_Motion_2}. A realistic SCM consists of a collection of 
open arrays of spins coupled via the exchange interaction. The length distribution of these arrays is determined by the spatial distribution of defects in a sample. 
In this sense, open boundary conditions give a more accurate description of SCM dynamics than periodic boundary conditions. 
However, we start considering periodic boundary conditions because calculations are less involved but still provide insight into the essential features which are not affected by the presence of defects. \\

When no external field is applied, $h_p=0$, the dependence on $r_{p,l}$ disappears from~\eqref{Glauber_eq_Motion_2} that then reduces to a linear system of first-order differential equations. The corresponding eigenvalue problem involves a circulant matrix and is diagonalized by a discrete Fourier transform. 
A general solution takes the form $s_{p}=\sum_q \tilde{s}_q\E^{iqp}\E^{-\lambda_q t}$, with 
\begin{equation} 
\label{lambda_q}
\lambda_q =\alpha\left(1-\gamma\cos q\right)
\end{equation} 
and $q=0, 2\pi/N,\dots, 2\pi(N-1)/N$ set by periodic boundary conditions. The initial configuration determines, instead, the Fourier amplitudes. If the system is prepared into a ferromagnetic saturated state with $s_{p}=1$ for every $p$, the only nonzero Fourier component corresponds to $q=0$, that is $\tilde{s}_0=1/N$. Accordingly, the magnetisation is expected to follow a mono-exponential relaxation with 
a characteristic time scale $\tau =1/\left[\alpha\left(1-\gamma\right)\right]$. For ferromagnetic coupling, $J>0$, $\tau$  diverges exponentially at 
low temperature like $\E^{4\kappa}$. Because of this divergence, some ferromagnetic ordering may persist over macroscopic time scales
in the absence of applied field. 
In this sense, the work of Glauber has foreseen what would be observed in SCMs about forty years later. The realization of these systems 
gave the opportunity to generalize the original Glauber model to include features of realistic SCMs and specific experiments. \\
The response to a tiny a.c. field $B=B'\E^{-i\omega t}$ is modelled by linearising the hyperbolic tangent in~\eqref{Glauber_eq_Motion_2}. 
But this does not eliminate the dependence on pair-spin correlations. Already Glauber circumvented this problem replacing $r_{p-1,p}$ and $r_{p,p+1}$
by their equilibrium average, equal to $\tanh(\kappa)$~\cite{Glauber}. Limiting himself to equivalent spins ($g_p=g$ independent of the site),   
he predicted that the a.c. susceptibility was of the form~\eqref{chi_ac}, provided that $\chi_\text{eq}$ was taken as the static susceptibility of the Ising model and 
$\tau =1/\left[\alpha\left(1-\gamma\right)\right]$. 
As mentioned in the previous section, non-collinearity among local anisotropy axes is more the rule rather than an exception. 
This affects Hamiltonian~\eqref{Ising_Hamiltonian_in_Glauber} through the site-dependent Land\'e factor.   
The spatial periodicity of $g_p$ defines the magnetic unit cell. It is worth remarking that the periodicity of $g_p$ 
generally depends on the direction along which the magnetic field is applied. The simplest case of a two-fold non-collinear Ising chain is sketched in Fig.~\ref{two-fold-chain}.  
With relatively small effort, an analytic formula for the a.c. susceptibility can be derived, which accounts for non-collinearity or non-equivalence of  
magnetic centres~\cite{JPCM_Vindigni_09}. For $\omega \ll \alpha$, a resonant behaviour, i.e. a frequency-dependent peak in $\chi(\omega)$, is expected only when the field is applied along specific crystallographic directions (e.g., the $\underline{c}$ axis in Fig.~\ref{two-fold-chain}).  
 In particular, those directions are the ones along which the ground-state magnetisation is uncompensated.     
This prediction for the dynamic response of non-collinear Ising chains was indeed supported by experiment~\cite{Eur_Phys_Lett,JPCM_Vindigni_09,JACS_Bernot}. 

For several years the truncation schemes summarized above had represented the starting point for generalizations of the Glauber model which aimed at giving a better account for the characteristics of real SCMs. 
Then, the restriction to zero-field a.c. susceptibility prevented from modelling the dependence of relaxation time on static applied field.     
A breakthrough was represented by the work of Coulon and co-workers~\cite{Coulon_PRB_07} who actualized the local-equilibrium approximation for pair-spin correlations proposed by 
Huang in the seventies~\cite{Huang_PRA_73}. 
Let us start from refreshing the main ideas of local-equilibrium approximation for the case of 
periodic boundary conditions, as treated by Huang. Equivalent magnetic moments, $g_p=g$, coupled ferromagnetically will be assumed.  
With these hypotheses, the single-spin averages are independent of the site at thermodynamic equilibrium and read   
\begin{equation} 
\label{Ising_eq_m}
m=\langle \sigma_p \rangle_\text{eq}=\frac{\sinh(h)}{\Delta^{1/2}}\qquad\qquad\text{with} \quad\Delta=\E^{-4\kappa} +\sinh^2 (h)\;,
\end{equation} 
where $\langle \dots \rangle_\text{eq}$ stands for equilibrium average, given by Boltzmann statistics. 
Translation invariance holds also for nearest-neighbour pair-spin correlations, of our interest, which are given by  
\begin{equation} 
\Gamma=\langle \sigma_{p+1}\sigma_p \rangle_\text{eq}=\frac{\sinh^2(h)}{\Delta} + \frac{\E^{-4\kappa}\left[\cosh(h)-\Delta^{1/2}\right]}{\,\Delta\,\,\left[\cosh(h)+\Delta^{1/2}\right]} \;.
\end{equation} 
By means of~\eqref{Ising_eq_m}, $\sinh(h)$ and $\cosh(h)$ appearing in the nearest-neighbour correlation can be expressed in terms of $m$ and $\E^{-4\kappa}$, which yields   
\begin{equation} 
\label{Ising_eq_Gamma}
\Gamma=1- \frac{2(1-m^2)}{1+\sqrt{m^2+(1-m^2)\E^{4\kappa}}} \;.
\end{equation} 
In two physically relevant situations translational invariance may be assumed for time-dependent spin averages, $s_p$, as well. 
The first one corresponds to having equal initial conditions for all spins: $s_p(0)=\mu$, with $-1\le \mu \le 1$. Since $g_p=g$ have been assumed, this initial condition is simply realized when a magnetic field (possibly zero) has been switched on far in the past ($t\rightarrow-\infty$) and changed to some different value at time $t=0$.      
The second situation is a typical a.c. susceptibility experiment, for which only the stationary response to a tiny external drift is relevant.  
In these two cases, time-dependent spin averages become independent of the site and the label $p$ can be dropped from the variables $s_p$ in~\eqref{Glauber_eq_Motion_2}. 
The local-equilibrium approximation consists in assuming that~\eqref{Ising_eq_Gamma}, which establishes a closed relation between equilibrium spin averages and nearest-neighbour correlations,  
holds true for time-dependent averages as well, namely out of equilibrium.   
Equation~\eqref{Glauber_eq_Motion_2}, thus, simplifies as 
\begin{equation} 
\label{Glauber_eq_Motion_unif}
\frac{1}{\alpha}\frac{\D s}{\D t}=-\left(1-\gamma\right)s 
+\left(1- \gamma\Gamma\right)\tanh(h)\;,
\end{equation}
where $\Gamma$ is given by~\eqref{Ising_eq_Gamma} with $m$ is replaced by $s$ (time-dependent average). Within the Glauber model, the local-equilibrium approximation is nothing but a trick 
to truncate the hierarchy of kinetic equations. The resulting equations of motion are generally non-linear, 
the non-linearity arising from $\Gamma[s]$. 
Fortunately enough, equation~\eqref{Glauber_eq_Motion_unif} can be solved analytically~\cite{Huang_PRA_73}.  More importantly, for 
$t\rightarrow\infty$ the  exact steady-state solution is recovered. 
For instance, a mean-field truncation scheme might alternatively be assumed, setting $\Gamma=s^2$, but this would not reproduce the exact steady-state solution. 
Note that local-equilibrium approximation does not require small applied fields. For what concerns SCMs, much interest relates to the study of linear departures from equilibrium. 
Following Pini and co-workers~\cite{MG_Pini_PRB_11}, let us split the field  into a static contribution of any intensity ($h_0=\mu_\text{B} B_0g/kT$) plus an oscillating field 
of much smaller intensity $B'$ and with frequency $\omega$: $h=h_0+h'\E^{-i\omega t}$. As a consequence,  $s$ is expected to deviate slightly from its equilibrium value, $m(T,B_0)$, and~\eqref{Glauber_eq_Motion_unif} can be linearised as follows: 
\begin{equation} 
\label{Glauber_eq_Motion_unif_lin}
\frac{1}{\alpha}\frac{\D\delta s}{\D t}=-\left(1-\gamma+2\gamma\tanh^2(h_0)\right)\delta s + 
\left(1-\tanh^2(h_0)\right)h'\E^{-i\omega t}\;,
\end{equation}
where $\delta s =s(t)-m$ and the fact that $\Gamma[s]\approx \Gamma[m] + \left(d\Gamma/dm\right)_\text{eq}\delta s$ with $\left(d\Gamma/dm\right)_\text{eq}=2\tanh(h_0)$ has been used. 
The stationary behaviour is obtained inserting the trial solution $\delta s=\widetilde{\delta s}\,\E^{-i\omega t}$ in~\eqref{Glauber_eq_Motion_unif_lin}, which yields the a.c. susceptibility. 
The resulting formula is equivalent to~\eqref{chi_ac} and $\chi_\text{eq}$ is the susceptibility that would be obtained by differentiating $m$ in~\eqref{Ising_eq_m} with respect to $B$. 
This matching is a direct consequence of the fact that local-equilibrium approximation provides the exact steady-state solution for $s$.  
The relaxation time is, instead, given by the inverse of the prefactor of $\delta s$ in~\eqref{Glauber_eq_Motion_unif_lin}:
\begin{equation} 
\label{Local-eq_tau}
\tau=\frac{1}{\alpha\left(1-\gamma+2\gamma\tanh^2(h_0)\right)}\;.
\end{equation}  
The Glauber relaxation time is recovered in the limit $ h_0\rightarrow 0$ and, as already pointed out, diverges exponentially upon lowering temperature. 
Note that the net effect of a static field is that of removing such a divergence, though the dependence of the relaxation time on $B_0$ is much less dramatic than on temperature. 

The Glauber model was extended to weakly interacting spin chains by Z\v{u}mer~\cite{Zumer}. 
Similarly to Scalapino~\cite{Scalapino_quasi_1d}, he treated the interchain interaction as a mean field, limiting his analysis to the critical region around the transition to a 3D ordered phase. 
Equation~\eqref{Local-eq_tau} may allow generalizing Z\v{u}mer's results to lower temperatures, away from the critical region. 
A joint theoretical and experimental investigation of this phenomenon would provide important information on the critical behaviour of SCMs~\cite{Coulon_PRL_09}.  
A realistic model should, however, take into account finite-size effects induced by the presence of defects.

\section{Glauber model for finite chains\label{FS_Glauber}}
Though it may sound somewhat technical, the study of finite-size effects have played a central role in theoretical and experimental characterization of SCMs.  
As a first step, open boundary conditions need to be considered, which makes the transition probability of extremal spins take the form
\begin{equation}
\label{Transition_Prob_Field_Glauber_F-S}
\begin{split}
w_{\sigma_{1}\rightarrow -\sigma_{1}}&=\frac{1}{2}\alpha \left[1-\eta\,\sigma_{1}\sigma_{2}\right]\left[ 1-  \sigma _{1}\tanh(h_1) \right]\\
w_{\sigma_{N}\rightarrow -\sigma_{N}}&=\frac{1}{2}\alpha\left[1-\eta\,\sigma_{N}\sigma_{N-1}\right]\left[ 1-  \sigma_{N}\tanh(h_N) \right]\;,
\end{split}
\end{equation}
with $\eta=\tanh( J/k T)$, obtained again from the detailed-balance condition. The kinetic equations for spin located at boundaries are modified accordingly: 
\begin{equation} 
\label{Glauber_eq_Motion_F-S}
\begin{split}
\frac{1}{\alpha}\frac{\D s_{1}}{\D t}&=-\left(s_{1}-\eta \,s_{2} \right) 
+\left(1-\eta\, \, r_{1,2} \right)\tanh(h_1)\\
\frac{1}{\alpha}\frac{\D s_{N}}{\D t}&=-\left(s_{N}-\eta \,s_{N-1} \right) 
+\left(1-\eta \, \, r_{N-1,N} \right)\tanh(h_N)\;.
\end{split}
\end{equation}
In the absence of external field, the characteristic time scales can be deduced by inserting the trial solution  $s_k=\left(\mathcal{A}_p \E^{i k q} +  \mathcal{A}_r \E^{-i k q} \right) \E^{-\lambda_q t}$ into system~\eqref{Glauber_eq_Motion_2} that still holds for bulk spins, with labels $2\le k \le N-1$. The relation between $\lambda_q$ and $q$ remains the same as in~\eqref{lambda_q} but 
the values taken by $q$ are different from the case of periodic boundary conditions. Due to the loss of translation invariance, both amplitudes  $\mathcal{A}_p$ and $\mathcal{A}_r$  must be considered.  A pair of equations for these amplitudes are obtained inserting the trial solution into~\eqref{Glauber_eq_Motion_F-S}, with $\lambda_q$ given by~\eqref{lambda_q}. 
For $B=0$, this is a homogeneous system that only admits the trivial solution $\mathcal{A}_p=\mathcal{A}_r=0$ unless the determinant 
of the coefficients of $\mathcal{A}_p$ and $\mathcal{A}_r$ is zero. 
By requiring this, the following implicit equation for the values of $q$ is obtained~\cite{Barma_FS,Luscombe}:
\begin{equation}
\label{tangent_F-S} \tan\left[(N-1)q \right]=-\frac{2 \hat{\xi} \tan  q}{1-\hat{\xi}^2 \tan^2 q}
\end{equation}
with $\hat{\xi} =\eta/(\gamma-\eta)$. 
The $q=0$ solution has to be rejected because it is independent of $N$ for every temperature, which is not physical.   
The remaining solutions will be labelled with $\nu$, i.e., $\lambda_\nu$. 
For ferromagnetic coupling, $J>0$, the eigenfrequency corresponding to the slowest time scale can be expanded for low temperatures to get~\cite{da_Silva}
\begin{equation}
\label{lambda_1_F-S}
\lambda_{1}=\frac{2\alpha}{N-1} \E^{-2\kappa} + \mathcal{O}\left( \E^{-4\kappa}\right) \;.
\end{equation}  
The previous expansion contributed significantly to understanding SCMs. 
From~\eqref{lambda_1_F-S} one expects the slowest degrees of freedom of the system to equilibrate with a relaxation time $\tau_{N}\sim N \E^{2\kappa}$. 
The fact that the energy barrier at the exponent is halved with respect to Glauber's result suggests that, at low temperature, relaxation is driven by nucleation of a DW from a boundary. 
At higher temperatures, the Glauber behaviour is recovered. This happens when the  correlation length becomes significantly smaller than $N$ and physics becomes independent of boundary conditions.  
Thus, in real systems, the relaxation time is expected to diverge like  $\E^{4\kappa}$ at high temperatures -- when $\xi$ is much smaller than the average distance among defects -- and like $\E^{2\kappa}$ at low temperatures.   
The experimental observation of such a crossover represented an important step in establishing that SCM behaviour could, indeed, be described properly in the framework of Glauber dynamics~\cite{Coulon_PRB_04,Review_CC}. When $\xi\gg N$, the first step of relaxation is analogous to the nucleation of a critical droplet to reverse the magnetisation in metallic nanowires or elongated nanoparticles~\cite{B_Braun_AdvPhys_2012}. Depending on geometrical characteristics of the sample, non-uniform magnetisation reversal may be favoured with respect to the standard N\'eel-Brown mechanism (uniform rotation). The latter is known to follow an Arrhenius law, $\tau\sim \E^{\Delta_\tau /k T}$, with an energy barrier proportional to the sample volume. To the leading order, the temperature dependence is of the Arrhenius type also in the case of non-uniform magnetisation reversal, but  $\Delta_\tau$  typically does not depend on the sample size. 
This fact directly originates from the local character of DW excitations which serve as nuclei to initiate magnetisation reversal (relaxation), 
both in metallic nanowires and in SCMs at low temperature.

After being nucleated at one boundary, a DW may reach the other end of the chain with probability $\sim 1/N$ by performing an unbiased random walk~\cite{ICA_Vindigni_08}. 
This is at the origin of the dependence on $N$ appearing in~\eqref{lambda_1_F-S} and, consequently, in $\tau_{N}$. 
When this is the main channel for relaxation, in real SCMs one would expect to observe a decrease of the pre-exponential factor of the relaxation time by increasing the number of defects (see~\eqref{activation_xi-tau}); the energy barrier of the Arrhenius law should, instead, remain constant: $\Delta_{\tau_N}=2J$. 
This trend was qualitatively confirmed in Co(hfac)$_2$(NITPhOMe) compounds in which part of the Co$^{2+}$ ions were substituted, 
in different amounts, by non-magnetic Zn$^{2+}$ atoms~\cite{PRL_Bogani,APL_Vindigni}.  
The fact that the pre-exponential factor increases with the system size is typical of a sizeable time elapsed during DW propagation in the relaxation process. 
In passing, we note that the opposite trend, i.e., a decrease of the pre-exponential factor of relaxation time with increasing the system size, 
was predicted for nanowires in which magnetisation reversal is forced to initiate 
from the bulk (e.g., in toroidal samples or with enhanced anisotropy at the ends)~\cite{B_Braun_AdvPhys_2012}. 
In that case, the probability to nucleate a soliton-antisoliton pair increases with $N$ and the reversal rate consequently.

The local-equilibrium approximation may also be used to decouple the hierarchy of Glauber equations when a finite field is applied to an open chain.  
It is convenient to linearise directly~\eqref{Glauber_eq_Motion_2} with respect to $\delta s_p= s_p - m_p$. Note that the equilibrium values $m_p$ are now 
site-dependent due to the lack of translation invariance. The kinetic equations for 
$\delta s_k$ contain the variation of nearest-neighbour correlation functions $\delta r_{p-1,p}$ and $\delta r_{p,p+1}$. 
For a chain of $N$ equivalent spins, with $g_p=g$,  Matsubara and co-workers provided a set of analytic relations to express 
equilibrium correlations $\langle \sigma_p \sigma_{p+1} \rangle_\text{eq}$ as functions of single-spin averages of open chains of different length~\cite{Matsubara_73}. 
If one assumes that such relations still hold true out of equilibrium, 
pair-spin variations can be written in terms single-spin averages: $\delta r_{p,p+1}=A_{N,p}\delta s_p + B_{N,p}\delta s_{p+1}$ 
with $A_{N,p}$ and $B_{N,p}$ depending only on equilibrium quantities (the reader is addressed to Ref.~\cite{Coulon_PRB_07} for details). 
With the same convention introduced in~\eqref{Glauber_eq_Motion_unif_lin} the response to an a.c. field $B'\E^{-i\omega t}$ superimposed to a static field $B_0$ is 
described by a system of linear equations of the form: 
\begin{equation} 
\label{Glauber_ac_F-S}
\frac{\D \boldsymbol\Sigma}{\D t}=-M \boldsymbol\Sigma + \alpha \left(1-\tanh^2(h_0)\right)h'\E^{-i\omega t}\, \boldsymbol\Psi\;,
\end{equation}     
where $\boldsymbol\Sigma=\left( \delta s_1,\dots, \delta s_N\right)^\text{T}$; the matrix $M$ and the vector $\boldsymbol\Psi$ are only functions of equilibrium averages,  
model parameters, temperature and static field (explicit expressions can be found in Ref.~\cite{MG_Pini_PRB_11}). 
Let $\boldsymbol\phi_\nu$ and $\lambda_\nu$ be the eigenvectors and eigenvalues of $M$, namely $M\boldsymbol\phi_\nu=\lambda_\nu\boldsymbol\phi_\nu$.  
The stationary solution of~\eqref{Glauber_ac_F-S} then reads 
\begin{equation} 
\label{Sigma_F-S}
\boldsymbol\Sigma=\alpha\left(1-\tanh^2(h_0)\right)h'\E^{-i\omega t}\sum_\nu  \frac{\boldsymbol\Psi \cdot \boldsymbol\phi_\nu}{\lambda_\nu}\frac{\boldsymbol\phi_\nu}{1-i\omega\tau_\nu}
\end{equation}    
with $\tau_\nu=1/\lambda_\nu$. 
The dynamic susceptibility is given by $\chi(\omega)$=$g\mu_\text{B} \E^{i\omega t}\sum_p  \delta s_p/B' $, where $\delta s_p$ are the components of the
$\boldsymbol\Sigma$ vector in~\eqref{Sigma_F-S}.  
With respect to the case with periodic boundary conditions, the choice of a site-independent Land\'e factor does not yield an a.c. response dependent on a single relaxation time. 
The relative weight of different contributions labelled by $\nu$ shall depend on temperature and on the static field $B_0$. 
In practice, the matrix $M$ can be diagonalized numerically for any values of $B_0$ and $T$. The size of this matrix, $N$ by $N$, is set by the number of spins in the chain. 
Realistic values of $N$ fall in the range $10-10^4$, meaning that $\chi(\omega)$ can easily be computed with standard diagonalisation routines.  
Among other things, this allows checking whether a unique relaxation time is dominating the summation~\eqref{Sigma_F-S} and thus $\chi(\omega)$.  
When the distribution of defects in a SCM compound is not peaked, an average over all the possible lengths may be required
to compare the theoretical susceptibility with experiments~\cite{APL_Vindigni,PRL_Bogani,MG_Pini_PRB_11}. 
Analogously, for a comparison with experiments on powder samples an average over all the possible orientations of the applied field with respect to the 
easy axis might be needed~\cite{Coulon_PRB_07}. Due to space limitations, we prefer not to enter the details of those averaging procedures but, rather, address to the existing literature.

The divergence of relaxation time upon lowering temperature can be interpreted as critical slowing down. The 1D Ising model displays a magnetic phase transition at zero temperature, meaning that the critical point is located at the origin of the $(T, B_0)$ plane, that is $T=0$ and $B_0=0$. 
Since the divergence of the  correlation length is hampered by defects, it is more appropriate to investigate the critical behaviour of SCMs with finite-size scaling. 
For $B_0=0$, Luscombe \textit{et al.} noted that the ratio between the relaxation time of a finite chain, $\tau_N$, and that of the infinite chain, $\tau$ originally obtained by Glauber, 
is a \textit{universal} function of $x=N/\xi$, 
when both $N$, $\xi\gg$1: 
\begin{equation} 
\label{Luscombe}
\frac{\tau_N}{\tau} =f(x)=\frac{1}{1+\left(\frac{\omega(x)}{x}\right)^2}\;,
\end{equation}     
where $\omega(x)$ is the smallest root of the transcendental equation $\omega\tan(\omega/2)=x$~\cite{Luscombe}. 
By definition, $f(x)$ tends to one for $x\gg1$; while for $x\ll1$ it is $f(x)\simeq x/2$. The reader may easily verify this limit by using formulae   
$\tau\simeq \E^{4\kappa}/2\alpha$, $\tau_N \simeq N \E^{2\kappa}/2\alpha$ and $\xi\simeq \E^{2\kappa}/2$, which hold for $N$, $\xi\gg$1 (see~\eqref{lambda_1_F-S} and~\eqref{lambda_q}). 
More recently, Glauber dynamics of the open chain in presence of realistic fields was studied by Coulon 
and co-workers~\cite{Coulon_PRB_07} who found 
\begin{equation} 
\label{Ratio_B}
\frac{\tau_N(B_0=0)}{\tau_N(B_0\ne0)} = 1 + a^2 h_0^2\;;
\end{equation} 
remarkably, the constant on the right-hand side is given by $a=2\xi f(x\sqrt{2/3})$, $f(x)$ being the scaling function defined in~\eqref{Luscombe}. 
For $x\gg1$, the limit $a=2\xi$  is recovered by expanding the hyperbolic tangent in~\eqref{Local-eq_tau} (remember that $f(x)\rightarrow 1$ in this limit). 
In the opposite limit, one has $a=\sqrt{2/3}N$ consistently with the work of Schwarz developed in the context of helix-coil transition of polypeptides~\cite{Schwarz_BioPol_68}. 
The quadratic dependence on $B_0$ of the ratio of relaxation times in the vicinity of the critical point stated by~\eqref{Ratio_B} 
was confirmed by experiment: first, in SCMs made up of repeating trinuclear units, 
Mn$^{3+}$-Fe$^{3+}$-Mn$^{3+}$ and Mn$^{3+}$-Ni$^{2+}$-Mn$^{3+}$~\cite{Coulon_PRB_07}, later in Co(hfac)$_2$(NITPhOMe) compounds~\cite{MG_Pini_PRB_11}.   
As pointed out in Ref.~\cite{Coulon_PRB_07}, the quadratic dependence on $B_0$ is also expected for SMMs. 
In fact, when repeating units in a spin chain consist of SMM-like centres an additional dependence on temperature and on $B_0$ enters the Glauber model through the attempt frequency $\alpha$. 
Thus, information about the 1D universality class is somehow contained in the scaling function $f(x)$ rather than in the quadratic take-off of 
$\tau_N(B_0\ne0)$ as a function of the applied field.  

In summary, the Glauber model prescribes that precise relations among characteristic energy scales shall hold for a text-book SCM. 
Recalling~\eqref{activation_xi-tau} and~\eqref{xi_chi_stat}, the barrier controlling the divergence of the correlation length can be directly deduced from static susceptibility measurements at high enough temperature. 
The last condition is required in order for $\xi$ to be smaller than the distance among defects. When this ceases to hold true, 
a saturation of the product  $\chi\, T$ is observed, at low $T$. 
According to the Ising model $\Delta_\xi=2J$, which implies that the energy barrier of the relaxation time is expected 
to be $\Delta_\tau=2\Delta_\xi=4J$ at high temperature and $\Delta_{\tau_N}=\Delta_\xi=2J$ at low temperature. 
Moreover, the crossover between the thermodynamic limit ($\xi\ll N$) and the finite-size regime ($\xi\gg N$) should be described by~\eqref{Luscombe} and~\eqref{Ratio_B} in the presence of a static applied field. 
SCMs represent a class of model systems in which most of these predictions were confirmed.  Often, finding a quantitative agreement required \textit{ad-hoc} generalizations of Glauber's idealized picture, without renouncing its basic concepts. Some of those generalizations will be discussed in the next section.

\section{Beyond the Glauber model\label{beyond_Glauber}}
An important generalisation of the Glauber model relates the temperature dependence of the parameter $\alpha$~\cite{Shen}. 
Introduced in~\eqref{Transition_Prob_Field_Glauber} for the single-spin transition probability, this parameter turns out to be the 
proportionality coefficient between the low-temperature expansions of the  correlation length and the relaxation time: $\tau=2\xi^2/\alpha$.  
In other words, for the time scales of interest, one has that $\alpha=4D_\text{s}$, with $D_\text{s}$ being the diffusion coefficient for thermally-driven DW motion (see~\eqref{xi-tau}). 
Given this equivalence, we will focus on $D_\text{s}$ henceforth.  
For explaining the experimental results of a SCM made of Mn$^{3+}$-Ni$^{2+}$-Mn$^{3+}$ repeating units it was proposed that $D_\text{s}\sim \E^{-\Delta_\text{A}/kT}$, where $\Delta_\text{A}$ was the global effective anisotropy energy of each unit~\cite{Coulon_PRB_04}. 
The relationship between the energy barriers of the correlation length and the relaxation time was adapted accordingly:  $\Delta_\tau=2\Delta_\xi + \Delta_\text{A}$. The last formula has been validated by experiments on a variety of SCMs with sharp DWs. In those cases, 
it was also found that the energy barrier of $\tau$ has to be modified consistently at low temperature, namely  $\Delta_{\tau_N}=\Delta_\xi + \Delta_\text{A} $.   

One minor remark is that the $\Delta_\text{A}$ contribution to the energy barrier of the relaxation time is justified only when some single-ion anisotropy is present. 
For instance, we have seen that Co$^{2+}$ in distorted octahedral environment can be assumed to behave as an effective spin one-half at low temperature.  
This assumption is not consistent with a finite $\Delta_\text{A}$ for SCMs based on Co$^{2+}$.  
More importantly, the picture appears more blurred for broad DWs. 
Let us refer again to Hamiltonian~\eqref{Heisenberg_chain_classic}. As mentioned in Sect.~\ref{origin_of_slow_dyn}, for $J>D$ 
the correlation length in units of DW width is a universal function of the temperature expressed in units of DW energy:  $\xi/ w = \Lambda\left(\mathcal{E}_\text{dw}/kT\right)$. 
It has been known since the eighties that a spin wave can propagate across a broad DW acquiring a phase shift%
\footnote{Recently, magnonic applications of DWs which exploit such a phase shift has been proposed in the context of metallic nanowires~\cite{SW-DW_phase-shift_magnonic}.}~\cite{Fogedby_book}.   
In order to conserve the total magnetisation at short time scales, the DW is left displaced after this scattering event~\cite{Yan-Wang_PRL_11}. 
When many of such events occur incoherently and involve thermalised spin waves, the resulting DW motion may be assimilated with that of a Brownian particle.   
Indeed, still in the eighties, it was shown that 
$D_\text{s}$ scales like the square of the  ratio $kT/\mathcal{E}_\text{dw}$ in the absence of damping~\cite{Theodorakopoulos-Weller} and linearly when some damping term is included~\cite{Salerno}. 
Dimensional analysis suggests to complete the latter result as 
\begin{equation}
\label{diffusion-coeff_broad_DW}
D_\text{s} \propto \frac{w^2}{\tau_\text{d}} \frac{kT}{\mathcal{E}_\text{dw}}\;,
\end{equation}
with $\tau_\text{d}$ being a characteristic time scale of the problem, associated with short-time dynamics.   
Equation~\eqref{diffusion-coeff_broad_DW} is consistent with recent numerical results reported in Ref.~\cite{Billoni_PRB}. In the same paper, the activated behaviour of $D_\text{s}$ expected for sharp DWs was recovered as well. 
A qualitative argument for the different temperature dependence of $D_\text{s}$ expected for sharp and broad DWs can be given starting from zero-temperature dynamics. 
In the continuum formalism one finds that a field of any intensity applied along the easy axis is able to move a broad DW~\cite{Fogedby_book,Walker_74,Enz,DW-motion-nanowires}.   
In the opposite limit, it was shown that a finite threshold field is needed to let a sharp DW propagate~\cite{Barbara}.    
In this case, translating a DW requires local modifications of the spin profile, which create an effective Peierls potential. 
This potential is periodic with respect to the position of the DW centre and the difference between its minima and maxima decreases exponentially with increasing the DW width~\cite{Hilzinger,Yan-Bauer_PRL_2012}, 
till it vanishes in the continuum limit. It seems, therefore, plausible to expect a thermally-activated diffusion coefficient \textit{only} for sharp DWs. 

While for sharp DWs the relaxation time depends on $J$, $D$ and $T$ independently, our present understanding of SCMs suggests that 
$\tau$ should depend only on the ratio $\mathcal{E}_\text{dw}/kT$ for broad DWs. This can be readily deduced by relating $\tau$ to the correlation length 
$\xi=w\Lambda\left(\mathcal{E}_\text{dw}/kT\right)$ by means of the random-walk argument and~\eqref{diffusion-coeff_broad_DW}
(remember that this argument holds only for $\xi\ll N$)~\cite{Billoni_PRB}. 

The standard theoretical framework to deal with magnetisation dynamics is the Landau-Lifshitz-Gilbert (LLG) equation. In that context, one expects  $\tau_\text{d}$ introduced in~\eqref{diffusion-coeff_broad_DW} to be of the order of the dumping time: 
$\tau_\text{d}\simeq (1+\alpha_\text{G}^2)/ (\alpha_\text{G} \gamma_0 H_\text{A})\simeq \hbar/(2D\alpha_\text{G}) $, where $\alpha_\text{G}\ll 1 $ is the Gilbert damping~\cite{Gilbert_2004}, $\gamma_0$ is the gyromagnetic factor and  $H_\text{A}$ the anisotropy field. For values of $D$ that are realistic for SCMs, $\hbar/D$ falls in the picosecond range while the damping constant is typically $\alpha_\text{G}$=10$^{-1}$--10$^{-4}$. As slow dynamics is  usually probed in SCMs at time scales larger than milliseconds, clearly it pertains to long-time behaviour in the language of LLG equation. Moreover, since physics of SCMs is dictated by thermal fluctuations, a stochastic noise should be included in numerical simulations~\cite{Nowak,Cheng06PRL,Stariolo07}. 
In spite of the enormous improvements experienced in computational capabilities~\cite{GPU_LLG}, performing a stochastic-dynamic simulation which covers a time window of several orders of magnitudes still remains prohibitive. 
In this sense, the brute-force approach to SCMs dynamics does not seem promising for the next future.

With respect to the sharp-wall case, there is no analogous of the Glauber's formalism for SCMs with broad DWs. 
Experimental realizations basically consist of ferrimagnetic chains alternating Mn$^{3+}$ with an organic radical~\cite{Large-DW_SCMs}. 
A reasonable model is the one which produces the kernel~\eqref{Mn$^{3+}$_kernel}, where Mn spins are treated as classical vectors.  
In the experimentally accessible region $\Delta_\xi$ can be much smaller than $\mathcal{E}_\text{dw}$ -- up to about half of it -- due to spin-waves renormalisation~\cite{Yamashita_RSC_2012}; 
while for sharp DWs one has $\Delta_\xi=\mathcal{E}_\text{dw}$ at any temperature $kT\!<\!J$~\cite{Billoni_PRB}.  
This fact needs to be taken into account in the experimental characterization of SCMs with broad DWs (see Fig.~\ref{fig_xi}).   
For what concerns the barrier of the relaxation time, the available experimental results yield $\Delta_\tau$ about 10--20\% times larger 
than $\mathcal{E}_\text{dw}$\footnote{For these mixed chains $\mathcal{E}_\text{dw}=2\sqrt{JD}$, with a factor 2 of difference
with respect to Hamiltonian~\eqref{Heisenberg_chain_classic}~\cite{Yamashita_RSC_2012}.}~\cite{Large-DW_SCMs,Yamashita_RSC_2012}: 
much smaller than twice the DW energy at $T$=0 as predicted by Glauber for the Ising model.           
Making a definitive statement about the origin of energy scales involved in dynamics is not possible yet. 
Defects probably affect the nucleation and diffusion of broad DWs differently with respect to the Ising limit. 
In metallic nanowires, for instance, defects act as pinning centres for DWs or vortices. In SCMs a similar phenomenon may induce a reduction of DW mobility, namely $D_\text{s}$. 
Another possibility is that DWs may preferentially be nucleated at defects because it is energetically favourable\footnote{Accommodating the DW centre onto a defect reduces the anisotropy energy and $\mathcal{E}_\text{dw}$ consequently.}.      
Only a thorough characterisation of SCMs with broad DWs in which the concentration of defects may be controlled could allow answering those questions. 
At the same time, such a study would provide important information about the joint effect of defects and thermal fluctuations. This would also be  
relevant for DW dynamics in metallic nanowires that are typically described by the very same classical Heisenberg Hamiltonian~\eqref{Heisenberg_chain_classic}~\cite{DW-motion-nanowires,current-induced_DW,Nowak_3,B_Braun_AdvPhys_2012,race-track-memory,Tatara,Yan-Bauer_PRL_2012}.

\section{Conclusion and perspectives\label{conclusions}}
The title of the review we wrote about five years ago was ``Single-chain magnets: where to from here?''~\cite{JMC_Bogani}. The idea was that of reviewing critically what had been done in the synthetic, experimental and theoretical fields. The analysis indicated that the hunt for high-temperature blocking magnets was going to continue. This has been confirmed but with the explosion of the interest for Lanthanides with the challenging  difficulties associated with the large unquenched orbital moment~\cite{Lanthanides_MolecularMag}. 
Much more work shall be done especially in theory. Another field which is developing fast is that of \textit{ab initio}, 
DFT calculations which are rapidly complementing/substituting Ligand-field approaches~\cite{PRB_Luzon,DFT}. 
Far-from-equilibrium dynamics and aging~\cite{Glassy-reviews,Prados} as well as the interplay between SCM behaviour and quantum effects~\cite{WW_PRL_05,Coldea} call for a more systematic investigation. 
The comparison of the properties of molecular nanomagnets with elongated magnetic nanoparticles and magnetic nanowires has been stated a few times throughout the chapter~\cite{B_Braun_AdvPhys_2012,Billoni_PRB}.
We feel that SCMs can provide good insight into the finite-temperature behaviour of such nanosystems. 
Finally, as a matter of facts, molecular systems have already entered the domains of spintronics~\cite{Molec_spintronics} and quantum computing~\cite{Quantum-Comp}. 
In future, besides their traditional role as model systems, SCMs can possibly find their place in those applicative research contexts.

\section*{Acknowledgements}
This chapter is the result of several collaborations and fruitful, sometimes animated discussions with many colleagues.  
For this valuable contribution we are sincerely grateful to 
R. Sessoli, M. G. Pini, A. Rettori, L. Bogani, R. Cl\'erac, C. Coulon, M. Verdaguer, J. Villain, V. Pianet, T. T. Michaels, B. Sangiorgio, G. Venturi, H. Miyasaka, W. Wernsdorfer, and O. V. Billoni. 
We would also like to thank L. Sorace, M. G. Pini, F. Totti and L. G. De Pietro for the precious help provided in the editing phase and their patient and careful reading of the manuscript.

%
%
%
%
%



\end{document}